\documentclass[10pt,aps,prc,twocolumn,amsmath,amssymb,superscriptaddress,floatfix]{revtex4-2}
\usepackage{graphicx}           % \includegraphics
\usepackage{float}              % [H] placement
\usepackage{amsmath}            % equation environments
\usepackage{physics}

\usepackage[T1]{fontenc}
\usepackage[utf8]{inputenc}

\usepackage{xcolor}             % \textcolor, \color

\usepackage{multirow}

\usepackage{hyperref}

\usepackage{listings}
\usepackage{tikz}
\usepackage[compat=1.1.0]{tikz-feynman}

\newcommand{\be}{\begin{equation}}
\newcommand{\ee}{\end{equation}}
\newcommand{\beq}{\begin{eqnarray}}
\newcommand{\eeq}{\end{eqnarray}}
\newcommand{\ba}{\begin{array}}
\newcommand{\ea}{\end{array}}

\begin{document}

\title{Amplitude-Based Analysis of QED Radiative Corrections \texorpdfstring{\\}{ } to Electroproduction of \texorpdfstring{$\eta$}{eta}-Mesons on Protons}

\date{\today}

%=============================================================
%--------------------- AUTHOR LIST ---------------------------
%==============================================================
\author{\mbox{Isabella~Illari}}
\altaffiliation{Corresponding author: \texttt{iti2103@gwmail.gwu.edu}}
\affiliation{Institute for Nuclear Studies, Department of Physics, The
    George Washington University, Washington, DC 20052, USA}

\author{\mbox{Andrei~Afanasev}}
\altaffiliation{E-mail: \texttt{afanas@gwu.edu}}
\affiliation{Institute for Nuclear Studies, Department of Physics, The
    George Washington University, Washington, DC 20052, USA}
    
\author{\mbox{William~J.~Briscoe}}
\altaffiliation{E-mail: \texttt{briscoe@gwu.edu}}
\affiliation{Institute for Nuclear Studies, Department of Physics, The
    George Washington University, Washington, DC 20052, USA}

\author{\mbox{Victor~L.~Kashevarov}}
\altaffiliation{E-mail: \texttt{kashevar@uni-mainz.de}}
\affiliation{Institut f\"ur Kernphysik, University of Mainz, D-55099 Mainz, Germany}

\author{\mbox{Axel~Schmidt}}
\altaffiliation{E-mail: \texttt{axelschmidt@gwu.edu}}
\affiliation{Institute for Nuclear Studies, Department of Physics, The
    George Washington University, Washington, DC 20052, USA}

\author{\mbox{Igor~I.~Strakovsky}}
\altaffiliation{E-mail: \texttt{igor@gwu.edu}}
\affiliation{Institute for Nuclear Studies, Department of Physics, The
    George Washington University, Washington, DC 20052, USA}
\noaffiliation

%--------------------------------------------------------------
%---------------------- ABSTRACT -----------------------------
%==============================================================
\begin{abstract}
A formalism for radiative correction calculations in exclusive $\eta$ electroproduction 
on the proton is presented, extending the treatment developed for the pion channel. The 
EXCLURAD code is used in the radiative correction procedure with EtaMAID-2023 multipole 
amplitudes. The cross-section correction factor $\delta$ varies by up to ${\sim}\,30\%$ 
across the resonance region $W = 1.49$--$2.0$~GeV at $E_{\rm beam} = 6.535$~GeV, with 
a local maximum near $W \simeq 1.66$~GeV driven by the $S_{11}(1535)$ and $S_{11}(1650)$ 
resonances. The beam-spin asymmetry is suppressed by 15--25\% at the same kinematics. 
Numerical results covering $Q^2 = 0.3$--$4.0$~GeV$^2$ and the full angular range are 
provided for kinematics relevant to CLAS12 experiments at Jefferson Lab.
\end{abstract}

\maketitle

%==============================================================
%----------------------- Introduction ------------------------
%==============================================================
\section{Introduction}\label{sec:intro} 
Understanding the electromagnetic transition amplitudes from the nucleon's ground state to excited states provides valuable insight into the nucleon's electromagnetic structure. Exclusive pseudoscalar meson electroproduction is one of the major sources of direct information about the spatial and spin structures of excited states. With the development of a high-intensity, high-duty-factor electron beam possessing a high degree of polarization, this field has reached a new level of quality. For the past several years, exclusive electroproduction of mesons has been the focus of extensive studies at various accelerator laboratories, including Jefferson Lab and MAMI.
 
The $\eta$ meson carries isospin $I=0$, so the $\eta N$ system is a pure $I=1/2$ state.  Only nucleon resonances ($N^{\ast}$) contribute to $\eta$ electroproduction; $\Delta$ states ($I=3/2$) are excluded entirely~\cite{Liu:1989eta, Ronchen:2015vfa}.  Pion channels offer no equivalent filter since $\pi N$ final states mix $I=1/2$ and $I=3/2$ components.  Near threshold, $\eta$ production is dominated by the $S_{11}(1535)$, whose $N\eta$ branching fraction of $30$--$55\%$ is comparable to its $N\pi$ decay~\cite{ParticleDataGroup:2024cfk}.  The $S_{11}(1650)$ and $P_{11}(1710)$ also couple appreciably to $N\eta$, and the recent upgrade of $N(1895)\,1/2^{-}$ to four-star PDG status~\cite{ParticleDataGroup:2024cfk} has added further interest to the channel.  These features have driven a sustained experimental program in $\eta$ electroproduction at Jefferson Lab with the CLAS and CLAS12 detectors~\cite{CLAS:2003vka, CLAS:2003qum, CLAS:2004ncx, CLAS:2005vxa, CLAS:2005ekq, CLAS:2006sjw, CLAS:2006ezq, CLAS:2007jpl, CLAS:2008ihz, CLAS:2012ich, CLAS:2012cna, CLAS:2012qga, CLAS:2014jpc, CLAS:2014fml, CLAS:2016qxj, CLAS:2017rgp, CLAS:2019cpp, CLAS:2022iqy, CLAS:2023akb}.  A general discussion of $\eta$ electroproduction and its role in baryon spectroscopy can be found in Ref.~\cite{Knochlein:1995qz} and the reviews cited therein.
 
The $\eta$ production threshold, $W_{\text{th}} = m_p + m_\eta \approx 1.486~\mathrm{GeV}$, lies only ${\sim}\,49~\mathrm{MeV}$ below the $S_{11}(1535)$ peak.  For comparison, the pion threshold sits ${\sim}\,159~\mathrm{MeV}$ below the $\Delta(1232)$.  This compressed gap limits the bremsstrahlung phase space near the dominant resonance: even moderate photon emission can shift the effective invariant mass below the $\eta$ threshold, producing a steeper and more structured radiative correction than in the pion channel. Accurate radiative corrections are therefore required to interpret the measured cross sections and polarization asymmetries.
 
A formalism for radiative-correction (RC) calculations in exclusive pion electroproduction on the proton was developed over 20~years ago~\cite{Afanasev:2002ee} and was later applied to pion electroproduction experiments at JLab and CERN (COMPASS).  A code EXCLURAD was developed for the RC procedure. In this paper, we extend EXCLURAD to the $\eta$ channel. The new contributions are: an explicit derivation of the leading-logarithm approximation for the $\eta$ channel, which does not appear in closed form in Ref.~\cite{Afanasev:2002ee}; an interface to EtaMAID-2023 multipole tables covering $W = 1.49$--$2.0$~GeV and $Q^2 = 0$--$5$~GeV$^2$; and a numerical output of ${\sim}\,265{,}000$ kinematic points made publicly available with an interactive browser-based explorer. The QED formalism is unchanged from Ref.~\cite{Afanasev:2002ee}. Two-photon exchange corrections were not included in the present analysis; they can be applied independently using \textit{e.g.}, a soft-photon approximation \cite{afanasev2013two, lee2025soft}. The numerical analysis is conducted in the kinematics of current experiments by the CLAS Collaboration~\cite{CLAS:2003vka, CLAS:2003qum, CLAS:2004ncx, CLAS:2005vxa, CLAS:2005ekq, CLAS:2006sjw, CLAS:2006ezq, CLAS:2007jpl, CLAS:2008ihz, CLAS:2012ich, CLAS:2012cna, CLAS:2012qga, CLAS:2014jpc, CLAS:2014fml, CLAS:2016qxj, CLAS:2017rgp, CLAS:2019cpp, CLAS:2022iqy, CLAS:2023akb}.
 
%====================================================================
%--------- Kinematics, observables, and conventions ----------------
%====================================================================
\section{Kinematics, observables, and conventions}\label{sec:kinobs}
We follow the kinematic conventions of EXCLURAD for exclusive electroproduction.
\begin{figure}[tbp]
    \centering
    \includegraphics[width=0.9\linewidth]{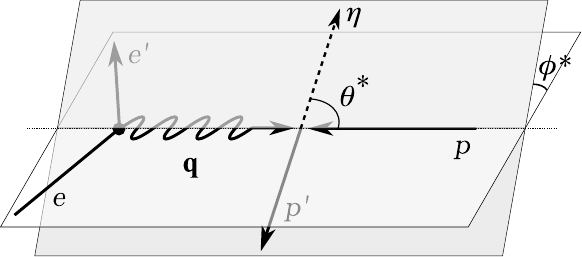}
    \caption{Kinematics of $\eta$ electroproduction in the center-of-mass frame, in the absence of any radiation, showing the angles $\theta^\ast$ and $\phi^{\ast}$.}
    \label{fig:kin}
\end{figure}
Four independent variables define the kinematics of an event. Labeling the incoming and outgoing electron 4-momenta as $k_1^\mu$ and $k_2^\mu$, these include the positive squared four-momentum transfer $Q^2 \equiv -(k_1^\mu-k_2^\mu)^2 > 0$, the invariant mass $W$ of the virtual photon–nucleon system, and the angles $\theta^{\ast}$ and $\phi^{\ast}$ of the outgoing meson momentum with respect to $\vb{q} = \vb{k}_1 - \vb{k}_2$ in the center-of-mass (c.m.) frame of the virtual photon and the initial nucleon. The azimuthal angle $\phi^{\ast}$ is defined relative to the electron scattering plane. The geometry is depicted in Fig.~\ref{fig:kin}.

As in Ref.~\cite{Afanasev:2002ee}, we consider the case where the outgoing electron and outgoing proton are detected while the outgoing meson is not detected. We define $\theta^\ast$ and $\phi^\ast$ based on the direction opposite to the momentum of the outgoing proton, following the convention used by EXCLURAD~\cite{Afanasev:2002ee} and adopted in CLAS and CLAS12 analyses. The undetected outgoing 4-momentum is
\begin{equation}
    \Lambda^\mu \equiv k_1^\mu + p^\mu - k_2^\mu - p^{\prime \mu} \>,
\end{equation}
where $p^\mu$ and $p^{\prime \mu}$ are the 4-momenta of the incoming and outgoing nucleons, respectively. This includes the 4-momentum of the outgoing meson and any 4-momentum carried by a radiated photon.

Additional discussion of the kinematic variables and the response-function decomposition of the Born cross section can be found in Secs.~II--III of Ref.~\cite{Afanasev:2002ee}.

In this work, we consider the radiative $\eta$ electroproduction reaction
\begin{equation}
    e~p \to e^\prime~p\;~\eta\;~(\gamma) \>,
\end{equation}
where $(\gamma)$ represents an undetected radiated photon and the meson takes the $\eta$-meson mass $m_\eta$. The production threshold is $W_\text{th} = m_p + m_\eta \approx 1.486~\mathrm{GeV}$.

The Lorentz-invariant inelasticity variable $v$ parametrizes radiative effects.
\begin{equation}
    v \equiv \Lambda^2 - m_\eta^2 \>,
\end{equation}
which vanishes at the Born point and increases with the energy of an unobserved bremsstrahlung photon. Experiments requiring exclusivity impose a cut $v < v_\text{cut}$ that controls how much of the radiative tail enters the exclusive sample. The value of $v_\text{cut}$ is typically chosen well below $v_{\text{max}}$, 
\begin{equation}
    v_\text{max} \equiv (W - m_p)^2 - m_\eta^2 \>.
\label{eq:vmax}
\end{equation}
This limit corresponds to the configuration in which all available energy has been radiated away, and the $\eta$ and proton are produced at rest in the c.m.\ frame, i.e., $\eta$ production at threshold. The value of $v_{\text{cut}}$ is chosen to exclude events in which an additional pion could have been produced in the undetected system. For the $\eta$ channel with a detected proton, the lightest additional hadron that could appear alongside the $\eta$ is a $\pi^0$, so the natural cut corresponds to the $\pi^0$ production threshold in the missing mass: $v_{\text{cut}} = (m_\eta + m_{\pi^0})^2 - m_\eta^2 = m_{\pi^0}^2 + 2\,m_\eta\,m_{\pi^0} \approx 0.166~\mathrm{GeV}^2$. Below this value, the only undetected particle kinematically allowed alongside the $\eta$ is a radiated photon; above it, the missing mass is large enough to accommodate an unobserved pion, and the event can no longer be treated as exclusive $\eta$ electroproduction. This is analogous to the logic applied in the pion channel in Ref.~\cite{Afanasev:2002ee}. When $v_{\text{cut}}$ exceeds $v_{\text{max}}$ at a given $W$ (which occurs for $W \lesssim 1.62$~GeV, where the compressed phase space near the $\eta$ threshold restricts $v_{\text{max}}$ below $v_{\text{cut}}$), the code uses $v_{\text{max}}$ as the upper integration limit; no extrapolation beyond the physical phase space boundary is performed. Any departures from $v_{\text{cut}} = 0.166~\mathrm{GeV}^2$ are stated explicitly (see, for example, the $v_{\text{cut}}$ scan in \hyperref[SM:S1]{SM~S1}).

We present results as two radiative-correction factors. For the unpolarized cross section,
\begin{equation}
    \delta(W,Q^2,\cos\theta^{\ast},\phi^{\ast}) \equiv \frac{d\sigma_{\text{obs}}}{d\sigma_0} \>,
\end{equation}
where $\sigma_{\text{obs}}$ includes radiative effects and $\sigma_0$ is the Born-level cross section; this matches Eq.~(75) of Ref.~\cite{Afanasev:2002ee}. For the beam spin asymmetry (BSA), $A_{LU}$, \textit{i.e.}, the asymmetry when scattering a longitudinally polarized electron beam from an unpolarized proton target,
\begin{equation}
    R_A(W,Q^2,\cos\theta^{\ast},\phi^{\ast}) \equiv \frac{A_{LU}^{\text{RC}}}{A_{LU}^{\text{Born}}} \>,
\end{equation}
where $A_{LU}^{\text{RC}}$ is the BSA including radiative effects, and $A_{LU}^{\text{Born}}$ is the Born-level BSA. Because $A_{LU}$ is a difference of cross sections, $R_A$ need not equal $\delta$; the two are distinct radiative-correction factors. 

%===================================================================
%----------- Updated EtaMAID amplitudes ----------------------------
%===================================================================
\section{Updated EtaMAID amplitudes}\label{sec:etamaid_update}
EtaMAID is an isobar model for $\eta$ photo- and electroproduction on the nucleon~\cite{Knochlein:1995qz, Chiang:2001as}. The model includes nucleon resonances in the $s$-channel parametrized by Breit-Wigner amplitudes, Born terms in the $s$- and $u$-channels, and vector-meson exchange in the $t$-channel. Because the $\eta$ meson carries isospin $I=0$, only nucleon resonances with $I=1/2$ contribute to the $\eta N$ final state, excluding all $\Delta$ ($I=3/2$) states that appear in pion electroproduction. This isospin filter provides a cleaner probe of the $N^\ast$ spectrum than the pion channel, where $I=1/2$ and $I=3/2$ contributions overlap.

The EtaMAID framework has undergone several revisions. The original model (EtaMAID2001/2003~\cite{Chiang:2001as}) described the non-resonant background with Born terms and $t$-channel $\rho$ and $\omega$ vector-meson exchanges and included eight nucleon resonances in Breit-Wigner form with partial waves up to $l=3$. In the EtaMAID2018 update~\cite{Tiator2018}, the background was replaced by a Regge-cut model incorporating $\rho$ and $\omega$ Regge trajectories with Pomeron and $f_2$ rescattering contributions. The resonance count was expanded from 8 to 21 $N^\ast$ states, partial waves were extended to $l=5$ (through $G_{17}$ and $G_{19}$), and energy-dependent unitarization phases were introduced for each resonance. Damping factors were applied to suppress double counting in the overlap region between Regge exchanges and $s$-channel resonances. The 2018 model was constrained by differential cross section and polarization data from MAMI, CLAS, GRAAL, and ELSA, covering $\eta$ and $\eta^\prime$ photoproduction on protons and neutrons up to $W \approx 4.5$~GeV~\cite{Tiator2018}.

For this work, new EtaMAID-2023 lookup tables of the electromagnetic multipole amplitudes were generated: magnetic ($M_{l\pm}$), electric ($E_{l\pm}$), and scalar/longitudinal ($S_{l\pm}$) transitions for orbital angular momentum $l = 0$ through $5$. The tables cover $W$ from threshold up to $6$~GeV and $Q^2$ up to $5~\mathrm{GeV}^2$. The present analysis uses only the region $W \leq 2$~GeV and $Q^2 \leq 5~\mathrm{GeV}^2$, matching the kinematic domain of CLAS12 $\eta$-electroproduction measurements at $E_{\rm beam} = 6.535$~GeV~\cite{Illari:2024lvw}; the broader coverage of the tables is intended to support future applications at higher beam energies. The EtaMAID interface is publicly available at \url{https://maid.kph.uni-mainz.de/eta2018/etamaid2018.html}, with multipole tables accessible at \url{https://maid.kph.uni-mainz.de/eta2018/eta-mult.html}. These tables serve as the hadronic model input to EXCLURAD, as described in the following section.

%==================================================================
%------ eta-channel implementation in EXCLURAD -------------------
%==================================================================
\section{\texorpdfstring{$\eta$}{eta}-channel implementation in EXCLURAD}\label{sec:eta_implementation}
The three technical additions relative to Ref.~\cite{Afanasev:2002ee} are documented in this section: the explicit leading-logarithm formula (Sec.~\ref{sec:ll}), the EtaMAID-2023 amplitude interface (Sec.~\ref{sec:etamaid_interface}), and the $\eta$-channel adaptation (Sec.~\ref{sec:eta_adapt}). EXCLURAD~\cite{Afanasev:2002ee} computes $\mathcal{O}(\alpha)$ QED radiative corrections to the five-fold differential cross section for $e~p \to e^\prime~p\,~\eta\,~(\gamma)$ using the Bardin-Shumeiko covariant IR-cancelation scheme. The formalism of Ref.~\cite{Afanasev:2002ee} is unchanged; this section documents the computational structure in enough detail to make the results in Sec.~\ref{sec:results} self-contained and derives the leading-logarithm approximation explicitly, as that expression does not appear in closed form in Ref.~\cite{Afanasev:2002ee}.

%==============================================================
%--- IV.A -----------------------------------------------------
%===============================================================
\subsection{Born cross section}\label{sec:born}
The Born-level five-fold differential cross section is decomposed as~\cite{Afanasev:2002ee}
\begin{equation}
\begin{split}
    \frac{d^5\sigma_0}{dW\,dQ^2\,d\Omega_\eta^\ast} = \Gamma_v \Bigl[&\sigma_T + \varepsilon\,\sigma_L + \varepsilon\,\sigma_{TT}\cos 2\phi^\ast \\
    &+ \sqrt{\tfrac{\varepsilon(\varepsilon+1)}{2}}\,\sigma_{LT}\cos\phi^\ast \\
    &+ h\sqrt{\tfrac{\varepsilon(1-\varepsilon)}{2}}\,\sigma_{LT^\prime}\sin\phi^\ast\Bigr] \>,
\end{split}
\label{eq:born}
\end{equation}
where $\Gamma_v$ is the virtual photon flux factor, $\varepsilon$ is the degree of transverse polarization of the virtual photon, $h = \pm 1$ is the longitudinal beam helicity, and $\phi^\ast$ is defined in Sec.~\ref{sec:kinobs}. The structure functions $\sigma_T$, $\sigma_L$, $\sigma_{TT}$, $\sigma_{LT}$, $\sigma_{LT^\prime}$ depend on $(Q^2, W, \cos\theta^\ast)$ and are supplied by EtaMAID-2023 as described in Sec.~\ref{sec:etamaid_interface}. The fifth structure function $\sigma_{LT^\prime}$ is the longitudinal-transverse interference response function associated with the beam helicity; the prime distinguishes it from the helicity-independent $\sigma_{LT}$. It is odd in $\phi^\ast$ and is the source of the beam-spin asymmetry $A_{LU}$.

%=================================================================
%--- IV.B --------------------------------------------------------
%=================================================================
\subsection{Radiative correction formalism}\label{sec:rc_formalism}
Let $s = (k_1^\mu + p^\mu)^2$ and $x = (k_2^\mu + p^\mu)^2$ denote the Bardin–Shumeiko Lorentz invariants, where $k_1^\mu$, $k_2^\mu$, and $p^\mu$ are the four-momenta of the incoming electron, outgoing electron, and target proton, respectively, as defined in Sec.~\ref{sec:kinobs}. 

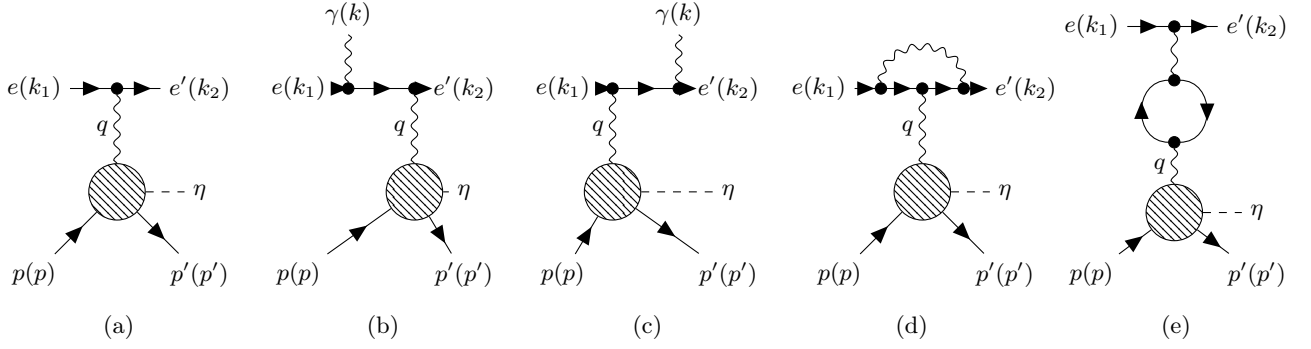
\begin{figure*}[tbp]
    \centering
    \begin{minipage}[b]{0.19\linewidth}
    \centering
    \begin{tikzpicture}[scale=0.55]
    \begin{feynman}
      \vertex              (i1) at (-2,   1)   {$e(k_1)$};
      \vertex [dot]        (v1) at ( 0,   1)   {};
      \vertex              (f1) at ( 2,   1)   {$e'(k_2)$};
      \vertex [blob]       (bl) at ( 0,  -1.5) {};
      \vertex              (i2) at (-2,  -3.5) {$p(p)$};
      \vertex              (f2) at ( 2,  -3.5) {$p'(p')$};
      \vertex              (f3) at ( 2,  -1.5) {$\eta$};
      \diagram* {
        (i1) -- [fermion]                (v1),
        (v1) -- [fermion]                (f1),
        (v1) -- [boson, edge label'=$q$] (bl),
        (i2) -- [fermion]                (bl),
        (bl) -- [fermion]                (f2),
        (bl) -- [scalar]                 (f3),
      };
    \end{feynman}
    \end{tikzpicture}
    \\[4pt] (a)
    \end{minipage}
    \begin{minipage}[b]{0.19\linewidth}
    \centering
    \begin{tikzpicture}[scale=0.55]
    \begin{feynman}
      \vertex              (i1)  at (-2,    1)   {$e(k_1)$};
      \vertex [dot]        (vr)  at (-0.8,  1)   {};
      \vertex [dot]        (v1)  at ( 0.8,  1)   {};
      \vertex              (f1)  at ( 2,    1)   {$e'(k_2)$};
      \vertex              (fph) at (-0.8,  2.8) {$\gamma(k)$};
      \vertex [blob]       (bl)  at ( 0.8, -1.5) {};
      \vertex              (i2)  at (-2,   -3.5) {$p(p)$};
      \vertex              (f2)  at ( 2,   -3.5) {$p'(p')$};
      \vertex              (f3)  at ( 2,   -1.5) {$\eta$};
      \diagram* {
        (i1)  -- [fermion]                (vr),
        (vr)  -- [fermion]                (v1),
        (v1)  -- [fermion]                (f1),
        (vr)  -- [boson]                  (fph),
        (v1)  -- [boson, edge label'=$q$] (bl),
        (i2)  -- [fermion]                (bl),
        (bl)  -- [fermion]                (f2),
        (bl)  -- [scalar]                 (f3),
      };
    \end{feynman}
    \end{tikzpicture}
    \\[4pt] (b)
    \end{minipage}
    \begin{minipage}[b]{0.19\linewidth}
    \centering
    \begin{tikzpicture}[scale=0.55]
    \begin{feynman}
      \vertex              (i1)  at (-2,    1)   {$e(k_1)$};
      \vertex [dot]        (v1)  at (-0.8,  1)   {};
      \vertex [dot]        (vr)  at ( 0.8,  1)   {};
      \vertex              (f1)  at ( 2,    1)   {$e'(k_2)$};
      \vertex              (fph) at ( 0.8,  2.8) {$\gamma(k)$};
      \vertex [blob]       (bl)  at (-0.8, -1.5) {};
      \vertex              (i2)  at (-2,   -3.5) {$p(p)$};
      \vertex              (f2)  at ( 2,   -3.5) {$p'(p')$};
      \vertex              (f3)  at ( 2,   -1.5) {$\eta$};
      \diagram* {
        (i1)  -- [fermion]                (v1),
        (v1)  -- [fermion]                (vr),
        (vr)  -- [fermion]                (f1),
        (vr)  -- [boson]                  (fph),
        (v1)  -- [boson, edge label'=$q$] (bl),
        (i2)  -- [fermion]                (bl),
        (bl)  -- [fermion]                (f2),
        (bl)  -- [scalar]                 (f3),
      };
    \end{feynman}
    \end{tikzpicture}
    \\[4pt] (c)
    \end{minipage}
    \begin{minipage}[b]{0.19\linewidth}
    \centering
    \begin{tikzpicture}[scale=0.55]
    \begin{feynman}
      \vertex              (i1) at (-2.5,  1)   {$e(k_1)$};
      \vertex [dot]        (vl) at (-1,    1)   {};
      \vertex [dot]        (vc) at ( 0,    1)   {};
      \vertex [dot]        (vr) at ( 1,    1)   {};
      \vertex              (f1) at ( 2.5,  1)   {$e'(k_2)$};
      \vertex [blob]       (bl) at ( 0,   -1.5) {};
      \vertex              (i2) at (-2,   -3.5) {$p(p)$};
      \vertex              (f2) at ( 2,   -3.5) {$p'(p')$};
      \vertex              (f3) at ( 2,   -1.5) {$\eta$};
      \diagram* {
        (i1) -- [fermion]                (vl),
        (vl) -- [fermion]                (vc),
        (vc) -- [fermion]                (vr),
        (vr) -- [fermion]                (f1),
        (vl) -- [boson, half left]       (vr),
        (vc) -- [boson, edge label'=$q$] (bl),
        (i2) -- [fermion]                (bl),
        (bl) -- [fermion]                (f2),
        (bl) -- [scalar]                 (f3),
      };
    \end{feynman}
    \end{tikzpicture}
    \\[4pt] (d)
    \end{minipage}
    \begin{minipage}[b]{0.19\linewidth}
    \centering
    \begin{tikzpicture}[scale=0.55]
    \begin{feynman}
      \vertex              (i1) at (-2,    1)   {$e(k_1)$};
      \vertex [dot]        (ve) at ( 0,    1)   {};
      \vertex              (f1) at ( 2,    1)   {$e'(k_2)$};
      \vertex [dot]        (vt) at ( 0,  -0.3) {};
      \vertex [dot]        (vb) at ( 0,  -1.8) {};
      \vertex [blob]       (bl) at ( 0,  -3.5) {};
      \vertex              (i2) at (-2,   -5)   {$p(p)$};
      \vertex              (f2) at ( 2,   -5)   {$p'(p')$};
      \vertex              (f3) at ( 2,  -3.5)  {$\eta$};
      \diagram* {
        (i1) -- [fermion]                (ve),
        (ve) -- [fermion]                (f1),
        (ve) -- [boson]                  (vt),
        (vt) -- [fermion, half left]     (vb),
        (vb) -- [fermion, half left]     (vt),
        (vb) -- [boson, edge label'=$q$] (bl),
        (i2) -- [fermion]                (bl),
        (bl) -- [fermion]                (f2),
        (bl) -- [scalar]                 (f3),
      };
    \end{feynman}
    \end{tikzpicture}
    \\[4pt] (e)
    \end{minipage}
    \caption{Feynman diagrams contributing to the Born and 
    $\mathcal{O}(\alpha)$ electroproduction cross sections.
    (a)~Born process; 
    (b)~initial-state bremsstrahlung off the incoming
    electron; 
    (c)~final-state bremsstrahlung off the outgoing electron;
    (d)~one-loop electron--photon vertex correction; and
    (e)~vacuum polarization insertion on the virtual photon propagator.
    The hatched blob represents all hadronic structure (EtaMAID-2023);
    $q = k_1 - k_2$ is the virtual photon four-momentum.
    Diagrams produced with Ti\textit{k}Z-Feynman~\cite{Ellis:2016jkw}.}
    \label{fig:feynman}
\end{figure*}

The five diagrams contributing to the Born and $\mathcal{O}(\alpha)$ corrected cross sections are shown in Fig.~\ref{fig:feynman}: the Born process~(a), initial- and final-state bremsstrahlung~(b,c), the one-loop vertex correction~(d), and the vacuum polarization insertion~(e).

The radiatively corrected cross section is (following Ref.~\cite{Afanasev:2002ee})
\begin{equation}
    \sigma_{\rm obs} = \sigma_0\,F_{\rm ext}\!\left(1 + \frac{\alpha}{\pi}\bigl[\delta_{VR} + \delta_{\rm vac}\bigr]\right) + \tau_{\rm hard} \>,
\label{eq:fullrc}
\end{equation}
where $F_{\rm ext}$ exponentiates the leading soft-photon IR contributions,
\begin{equation}
    F_{\rm ext} = \exp\!\left[\frac{\alpha}{\pi}(L_e - 1)\ln\frac{v_{\rm max}^2}{\hat{s}\,\hat{x}}\right] \>, 
\label{eq:fext}
\end{equation}
with $L_e = \ln\frac{Q^2}{m_e^2}$, and with 
Born-level invariants 
\begin{align}
    \hat{s} =& s - Q^2 - \tilde{v}_1^{(0)} \>,\\
    \hat{x} =& x + Q^2 - \tilde{v}_2^{(0)} \>,
\end{align}
(see Sec.~\ref{sec:ll} for the definition of $\tilde{v}_{1,2}^{(0)}$); $\delta_{VR}$ is the virtual-plus-soft finite remainder, given by
\begin{equation}
    \delta_{VR} = \tfrac{3}{2}L_e - 2 - \tfrac{1}{2}\ln^2\!\frac{\hat{x}}{\hat{s}} + \mathrm{Li}_2\!\!\left(1 - \frac{m_p^2 Q^2}{\hat{s}\,\hat{x}}\right) - \frac{\pi^2}{6} \>,
\label{eq:deltavr}
\end{equation}
and $\delta_{\rm vac}$ is the leptonic vacuum-polarization correction, summed over $e$, $\mu$, $\tau$, and light hadronic states.

The hard-bremsstrahlung contribution is the three-dimensional integral,
\begin{equation}
\begin{split}
    \tau_{\rm hard} = -\frac{\alpha^3}{512\pi^4 s^2 W^4} \int_0^{v_{\rm max}}\!\!dv &\int_{-1}^{1}\!\!d\cos\theta_\gamma \int_0^{2\pi}\!\!d\phi_\gamma \\
    &\times\mathcal{F}(v,\cos\theta_\gamma,\phi_\gamma) \>,
\end{split}
\label{eq:tauhard}
\end{equation}
where $\mathcal{F}$ is the squared radiative amplitude contracted with the hadronic tensor, with the hadronic model evaluated at the kinematically shifted arguments $(\tilde{Q}^2, \tilde{W}^2, \tilde{t})$ at each phase-space point.  The Bardin-Shumeiko subtraction renders $\mathcal{F}$ finite as $v \to 0$, so no regularization of the lower limit is needed.

The angular integrand has two sources of near-singular behavior: collinear singularities when the radiated photon is emitted parallel to $k_1$ or $k_2$ ($\cos\theta_\gamma = \cos\theta_1$ or $\cos\theta_2$ in the hadronic c.m.\ frame), and forward singularities near $\phi_\gamma = 0$ and $2\pi$. The code partitions the $(\cos\theta_\gamma, \phi_\gamma)$ domain into $5 \times 3 = 15$ sub-cells that isolate these peaks and applies adaptive quadrature to each sub-cell independently; a schematic of this partition is given in \hyperref[SM:S4]{SM~S4.1} (Figure ~\ref{fig:phasespace}). In the leading-log approximation (Sec.~\ref{sec:ll}), the $\cos\theta_\gamma$ and $\phi_\gamma$ integrals are performed analytically, reducing Eq.~(\ref{eq:tauhard}) to a one-dimensional integral over $v$. 

%==============================================================
%--- IV.C -----------------------------------------------------
%==============================================================
\subsection{Leading-logarithm approximation}\label{sec:ll}
When only the terms enhanced by $L_e = \ln(Q^2/m_e^2)$ are retained, the three-dimensional integral in Eq.~(\ref{eq:tauhard}) collapses to a one-dimensional integral over $v$. The formula implemented in the code is
\begin{equation}
    \tau_{\rm LL} = \frac{\alpha}{2\pi}(L_e - 1) \int_0^{v_{\rm max}} dv\; \Bigl[\mathcal{I}_{\rm ISR}(v) + \mathcal{I}_{\rm FSR}(v)\Bigr] \>,
\label{eq:taull}
\end{equation}
where the initial- and final-state radiation contributions are
\begin{align}
    \mathcal{I}_{\rm ISR}(v) &= \frac{(1+z_1^2)\,c_s\,\sigma_{\rm Born}^{(s)} - 2\,\tilde{\sigma}_0}{\hat{s}(v)\,(1-z_1)} \>,
\label{eq:isr} \\[4pt]
    \mathcal{I}_{\rm FSR}(v) &= \frac{\bigl[(1+z_2^2)\,c_x\,\sigma_{\rm Born}^{(x)} - 2\,\tilde{\sigma}_0\bigr]\,z_2}{\hat{x}(v)\,(1-z_2)} \>.
\label{eq:fsr}
\end{align}
To define the remaining quantities, consider the $\eta$ four-momentum shifted to inelasticity $v$, while its CM direction $(\cos\theta^\ast, \phi^\ast)$ is held fixed at the Born value. The shifted energy and momentum magnitude are
\begin{equation}
    \tilde{E}_\eta(v) = \frac{W^2 + m_\eta^2 - m_p^2 - v}{2W}, \qquad |\tilde{\vec{p}}_\eta(v)| = \frac{\tilde\lambda^{1/2}(v)}{2W} \>,
\label{eq:etashift}
\end{equation}
where
\begin{equation}
    \tilde\lambda^{1/2}(v) = \sqrt{\bigl(W^2 + m_\eta^2 - m_p^2 - v\bigr)^2 - 4W^2 m_\eta^2} \>.
\label{eq:lamsv}
\end{equation}
At $v = 0$, this reduces to the Born $\eta$ kinematics, while $v = v_{\rm max}$ corresponds to the threshold. The lepton-$\eta$ invariants at shifted kinematics are $\tilde{v}_1(v) = 2k_1\cdot\tilde{p}_\eta(v)$ and $\tilde{v}_2(v) = 2k_2\cdot\tilde{p}_\eta(v)$, with $\tilde{v}_{1,2}^{(0)} \equiv \tilde{v}_{1,2}(0)$ being the Born values appearing in 
Eqs.~(\ref{eq:fext})--(\ref{eq:deltavr}). The $v$-dependent kinematic denominators are
\begin{equation}
    \hat{s}(v) = s - Q^2 - \tilde{v}_1(v), \qquad \hat{x}(v) = x + Q^2 - \tilde{v}_2(v) \>,
\label{eq:sxhat}
\end{equation}
and the momentum fractions are
\begin{equation}
    z_1 = 1 - \frac{v}{\hat{s}(v)}, \qquad z_2 = \frac{\hat{x}(v)}{\hat{x}(v) + v} \>,
\label{eq:z12}
\end{equation}
so that $\hat{s}(1-z_1) = \hat{x}(1-z_2)/z_2 = v$, the standard Altarelli–Parisi variable. The hadronic invariant masses at shifted kinematics are
\begin{equation}
    W_s^2 = z_1 s - x - z_1 Q^2 + m_p^2, \qquad W_x^2 = s - \frac{x}{z_2} - \frac{Q^2}{z_2} + m_p^2 \>,
\label{eq:wshift}
\end{equation}
and the Born cross section is evaluated at two sets of shifted arguments,
\begin{align}
    \sigma_{\rm Born}^{(s)} &= \sigma_0\bigl(z_1 s,\; x,\; z_1 Q^2;\; z_1\tilde{v}_1,\; \tilde{v}_2\bigr) \>,
\label{eq:sigmas} \\[2pt]
    \sigma_{\rm Born}^{(x)} &= \sigma_0\!\left(s,\; \frac{x}{z_2},\; \frac{Q^2}{z_2};\; \tilde{v}_1,\; \frac{\tilde{v}_2}{z_2}\right) \>,
\label{eq:sigmax}
\end{align}
for initial- and final-state radiation, respectively. The subtraction term $\tilde{\sigma}_0 = \sigma_0(s, x, Q^2;\, \tilde{v}_1, \tilde{v}_2)$ is the Born cross section at physical invariants $(s, x, Q^2)$ but with the $\eta$ momentum at inelasticity $v$, so that $\tilde{\sigma}_0 \to \sigma_0$ as $v \to 0$. This ensures the integrand vanishes at the lower limit: as $v \to 0$, $z_{1,2} \to 1$, and $c_{s,x} \to 1$, both numerators in Eqs.~(\ref{eq:isr})--(\ref{eq:fsr}) vanish as $(1 + 1^2)\sigma_0 - 2\sigma_0 = 0$.

The factors $c_s$ and $c_x$ correct for the change in hadronic phase space 
when the invariant mass shifts from $W^2$ to $W_s^2$ or $W_x^2$. The Born 
cross section carries the $\eta$ CM momentum $\lambda^{1/2}(W^{\prime
 2})/(2W^\prime)$ as a flux factor in the numerator and $W^{\prime 2}$ in the 
denominator (see Eq.~(\ref{eq:born}) via Sec.~\ref{sec:etamaid_interface}). 
Correcting for both gives
\begin{equation}
    c_s = \frac{\tilde\lambda^{1/2}(v)}{\lambda^{1/2}(W_s^2)}\cdot\frac{W_s^2}{W^2}, \qquad c_x = \frac{\tilde\lambda^{1/2}(v)}{\lambda^{1/2}(W_x^2)}\cdot\frac{W_x^2}{W^2}
    \>,
\label{eq:cjac}
\end{equation}
where
\begin{equation}
    \lambda^{1/2}(W_{s,x}^2) = \sqrt{\bigl(W_{s,x}^2 + m_\eta^2 - m_p^2\bigr)^2 - 4W_{s,x}^2 m_\eta^2} \>.
\label{eq:lamsx}
\end{equation}
At $v = 0$, $\tilde\lambda^{1/2}(0) = \lambda^{1/2}(W^2)$, and $W_{s,x}^2 = W^2$, so $c_{s,x} = 1$ as required. The factors $c_s$ and $c_x$ correspond to $C_s$ and $C_x$ in Eq.~(59) of Ref.~\cite{Afanasev:2002ee}, which correct the hadronic phase-space flux when the invariant mass shifts from $W^2$ to $W_s^2$ or $W_x^2$ under photon emission. Reference~\cite{Afanasev:2002ee} writes the numerator using the Born $\eta$ momentum $\lambda^{1/2}(W^2)$; the form given here in Eq.~(\ref{eq:cjac}) instead uses $\tilde{\lambda}^{1/2}(v)$, the $\eta$ momentum evaluated at inelasticity $v$, making the $v$-dependence of the correction explicit. The two forms are identical at $v = 0$ and differ only in how the intermediate $v$-dependence is distributed between the numerator and the Born cross section $\tilde{\sigma}_0$. Both forms are implemented in the \textsc{exclurad} source code~\cite{exclurad_zenodo} and yield identical numerical results.

Equation~(\ref{eq:taull}) is evaluated by adaptive Simpson quadrature on 150 subintervals. Over the kinematic range in Sec.~\ref{sec:results}, the LL approximation deviates from the exact result by 2.7--8.8\% while running approximately 600 times faster.

%================================================================
%--- IV.D -------------------------------------------------------
%=================================================================
\subsection{Hadronic amplitude interface}\label{sec:etamaid_interface}
\begin{figure}[tbp]
    \centering
    \includegraphics[width=\linewidth,height=0.85\textheight,keepaspectratio]{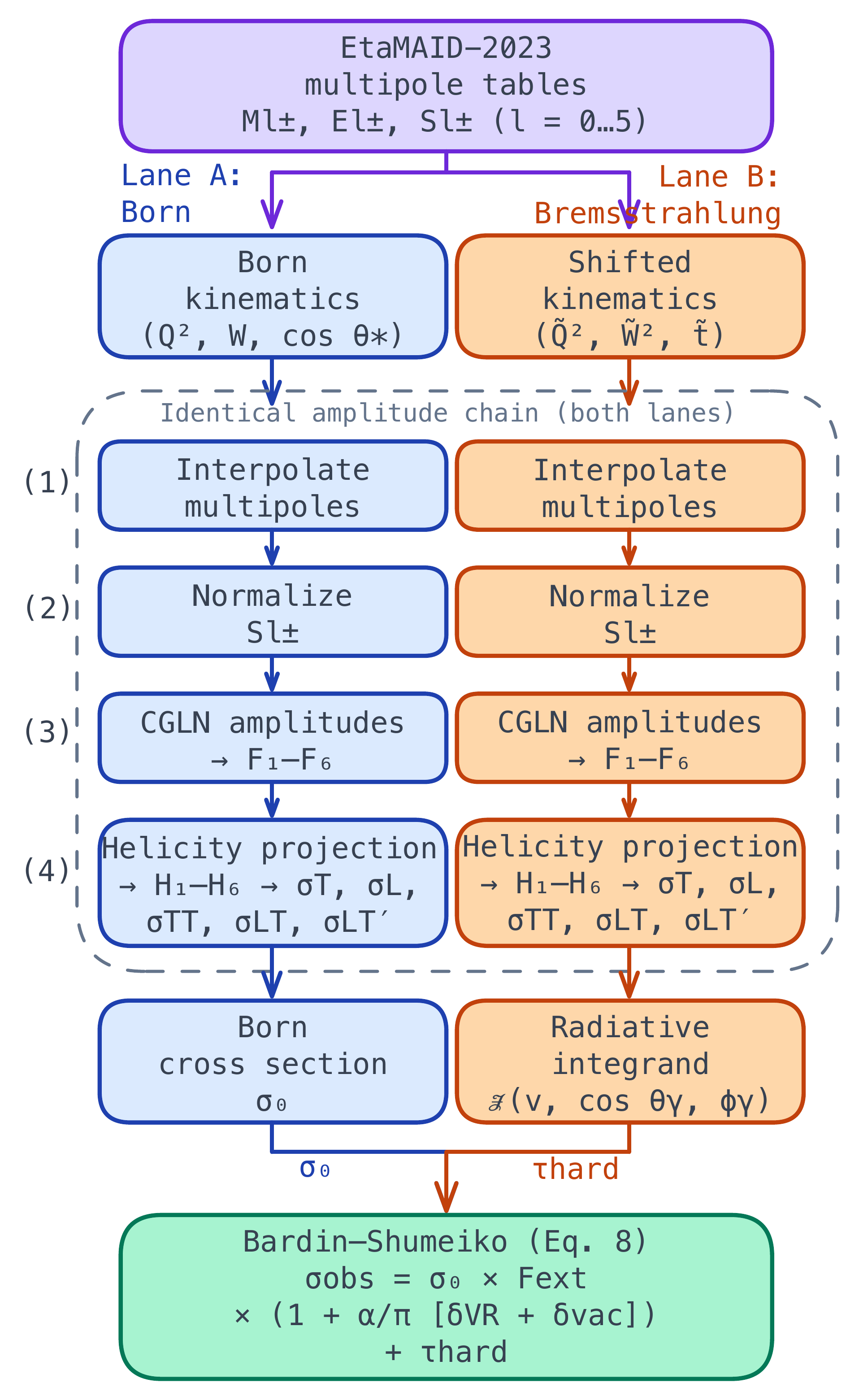}
    \caption{From multipoles to $\sigma_{\rm obs}$: the EtaMAID-2023 amplitude chain and its role in the radiative-correction calculation. The EtaMAID-2023 multipole tables (top, purple) store the electromagnetic multipole amplitudes $M_{l\pm}$, $E_{l\pm}$, $S_{l\pm}$ for $l = 0$--$5$ on a $(Q^2, W)$ grid. The same four-step chain (dashed boundary) converts these multipoles into structure functions in two parallel lanes. Lane~A (blue) runs at Born kinematics $(Q^2, W, \cos\theta^\ast)$ and produces $\sigma_0$; Lane~B (orange) runs at the radiatively shifted kinematics $(\tilde{Q}^2, \tilde{W}^2, \tilde{t})$ for each quadrature point in the bremsstrahlung integral and produces the radiative integrand $\mathcal{F}$. The four shared steps are: (1)~interpolation of the multipole amplitudes to the specified $(Q^2, W)$; (2)~normalization, including the $S_{l\pm} \to S_{l\pm} \times \nu_{\rm cm}/q_{\rm cm}$ rescaling and unit conversion~\cite{Tiator2018}; (3)~CGLN partial-wave summation to obtain $F_1$--$F_4$ from transverse multipoles and $F_5$, $F_6$ from scalar multipoles~\cite{Knochlein:1995qz}; and (4)~projection onto helicity amplitudes $H_1$--$H_6$, then construction of $\sigma_T$, $\sigma_L$, $\sigma_{TT}$, $\sigma_{LT}$, $\sigma_{LT^\prime}$. The two outputs combine at the bottom via the Bardin-Shumeiko formula (Eq.~(\ref{eq:fullrc})): $\sigma_0$ enters as the Born prefactor and $\mathcal{F}$ enters through $\tau_{\rm hard}$, yielding $\sigma_{\rm obs}$.}
    \label{fig:etamaid}
\end{figure}

The five structure functions in Eq.~(\ref{eq:born}) are built from EtaMAID-2023 multipole amplitudes $\{M_{l\pm}, E_{l\pm}, S_{l\pm}\}$ for partial waves $l = 0, \ldots, 5$, tabulated on a discrete $(Q^2, W)$ grid. At each grid node, the table stores the real and imaginary parts of all six amplitude types as (Re, Im) pairs, in units of $10^{-3}/m_{\pi^+}$~\cite{Tiator2018}; combinations forbidden by angular-momentum selection rules are set to zero. Intermediate grid points are obtained by bilinear interpolation in $(Q^2, W)$ with no extrapolation beyond the grid boundary. The grid covers $W = 1.4856~\mathrm{GeV}$ and $1.490$--$2.000~\mathrm{GeV}$ in $5~\mathrm{MeV}$ steps, $Q^2 = 0.00$--$5.01~\mathrm{GeV}^2$ in $0.01~\mathrm{GeV}^2$ steps. Where the public EtaMAID web interface is limited (nominally $W \leq 2~\mathrm{GeV}$), extended tables from the EtaMAID-2023 update described in Sec.~\ref{sec:etamaid_update} were used.
The amplitude chain that converts multipoles into structure functions is summarized in Fig.~\ref{fig:etamaid}. It proceeds in four steps. (1)~The multipole amplitudes are interpolated to the desired $(Q^2, W)$ point. (2)~The longitudinal multipoles $S_{l\pm}$ are rescaled by $\nu_{\rm cm}/q_{\rm cm}$ to convert from the $L$-multipole convention used in the tables to the $S$-multipole convention used in the cross-section formulas~\cite{Knochlein:1995qz}, and all amplitudes are converted from $10^{-3}/m_{\pi^+}$ to $\sqrt{\mu\mathrm{b}}$. (3)~A partial-wave sum constructs the CGLN amplitudes~\cite{ChewGoldberger1957} as functions of $\cos\theta^\ast$, parametrizing the hadronic electromagnetic current for pseudoscalar meson electroproduction off a spin-$1/2$ target~\cite{Knochlein:1995qz}. For real photons, the decomposition yields four amplitudes $F_1$--$F_4$, built from the transverse multipoles $M_{l\pm}$ and $E_{l\pm}$; for virtual photons, two additional amplitudes $F_5$ and $F_6$ enter through the longitudinal coupling, constructed from the scalar multipoles $S_{l\pm}$~\cite{Knochlein:1995qz}. (4)~The six CGLN amplitudes are projected onto helicity amplitudes $H_1$--$H_6$~\cite{Knochlein:1995qz}, from which the five structure functions $\sigma_T$, $\sigma_L$, $\sigma_{TT}$, $\sigma_{LT}$, and $\sigma_{LT^\prime}$ are constructed. Steps~(2)--(4) are implemented in the \textsc{exclurad} source code~\cite{exclurad_zenodo}.

This chain runs in two kinematic contexts (Fig.~\ref{fig:etamaid}, parallel lanes). Lane~A evaluates it at the Born kinematics $(Q^2, W, \cos\theta^\ast)$ to produce the Born cross section $\sigma_0$. Lane~B evaluates the same chain at the radiatively shifted kinematics $(\tilde{Q}^2, \tilde{W}^2, \tilde{t})$ at every quadrature point in the bremsstrahlung integral over $(v, \cos\theta_\gamma, \phi_\gamma)$, producing the radiative integrand $\mathcal{F}$. Steps~1--4 are identical between lanes; only the input kinematics differ. The two outputs feed the Bardin-Shumeiko formula, Eq.~(\ref{eq:fullrc}): $\sigma_0$ enters as the prefactor and $\mathcal{F}$ enters through $\tau_{\rm hard}$.
The QED formalism in Eqs.~(\ref{eq:fullrc})--(\ref{eq:cjac}) is independent of this hadronic chain; the hadronic model is called at kinematically shifted arguments inside $\tau_{\rm hard}$ and $\tau_{\rm LL}$, but the photon-emission kinematics and IR structure are those of Ref.~\cite{Afanasev:2002ee}.

%================================================================
%--- IV.E -------------------------------------------------------
%=================================================================
\subsection{Adaptation to the \texorpdfstring{$\eta$}{eta} channel}
\label{sec:eta_adapt}
Two changes were made to adapt EXCLURAD to the $\eta$-meson channel. First, the outgoing-hadron mass was set to $m_\eta = 0.54786~\mathrm{GeV}$, which places the production threshold at $W_{\rm th} \simeq 1.486~\mathrm{GeV}$ and sets $v_{\rm max}$ via 
Eq.~(\ref{eq:vmax}). Second, the hadronic model input was replaced by the EtaMAID-2023 amplitude tables; because the table format is identical to the existing MAID pion tables, no changes to the table-reading or interpolation routines were needed. The multidimensional integration tolerances for $\tau_{\rm hard}$ were tightened to a relative precision of $10^{-6}$ to improve convergence near the threshold. No changes were made to the QED formalism of Eqs.~(\ref{eq:fullrc})--(\ref{eq:cjac}). All results in Sec.~\ref{sec:results} use the exact $\mathcal{O}(\alpha)$ mode, which includes soft and hard bremsstrahlung, vertex and vacuum-polarization corrections, and soft-photon exponentiation. Unless otherwise noted, all figures use $v_{\rm cut} = 
0.166~\mathrm{GeV}^2$; individual figure captions state $E_{\rm beam}$ and any departures from this default. Convergence of $\tau_{\rm hard}$ was verified by comparing exact and leading-log results at representative kinematic points (\hyperref[SM:S4]{SM~S4.1}); the relative integration tolerance of $10^{-6}$ is sufficient that numerical integration error does not contribute meaningfully to any reported result. Full implementation details are in the repository (\hyperref[SM:S4]{SM~S4}).

%==================================================================
%-------------- Radiative-correction results ---------------------
%==================================================================
\section{Radiative-correction results}\label{sec:results}
All numerical results in this section use a beam energy of $E_{\text{beam}} = 6.535$~GeV, corresponding to the CLAS12 RG-K running period~\cite{Illari:2024lvw}; for comparison, Ref.~\cite{Afanasev:2002ee} used $E_{\text{beam}} = 1.645$~GeV for the pion channel.

%==================================================================
\subsection{\texorpdfstring{$W$}{W}-dependence of the correction factors}\label{sec:results_W}
\begin{figure}[tbp]
  \centering
  \includegraphics[width=\linewidth]{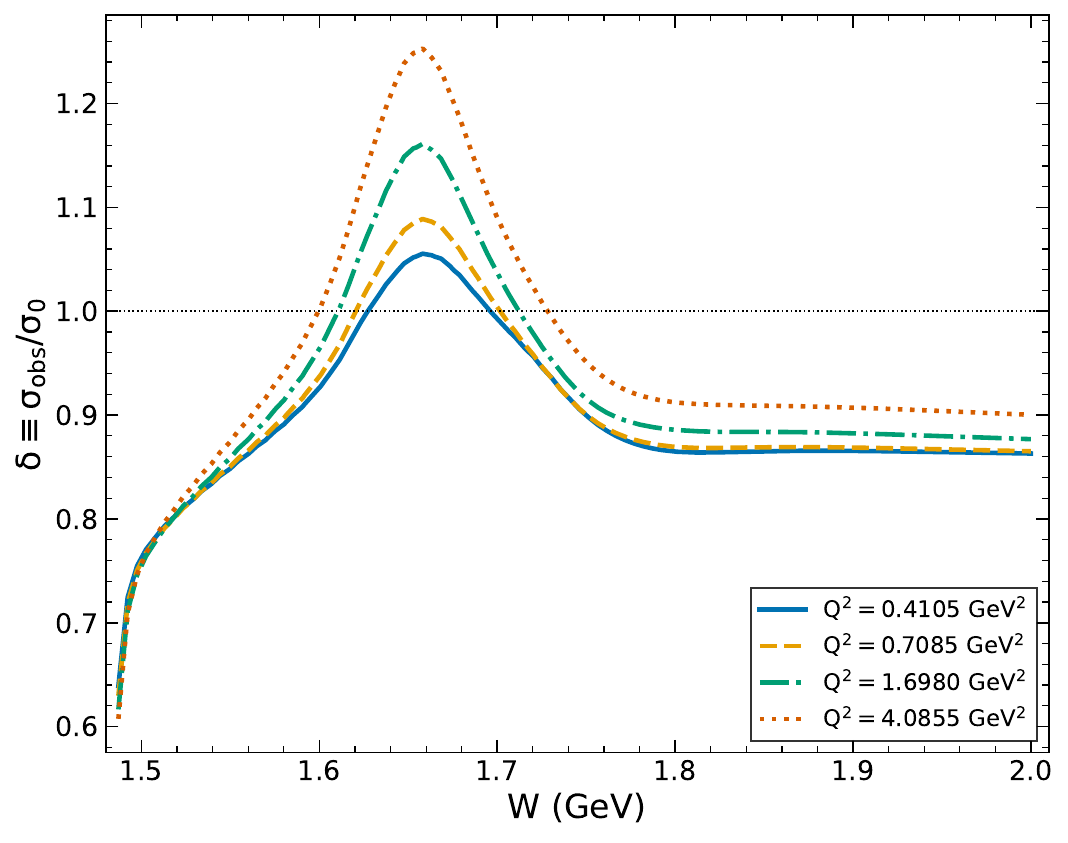}
  \caption{Cross-section RC factor $\delta \equiv \sigma_{\rm obs}/\sigma_{0}$ vs.\ $W$ at $\cos\theta^{\ast} = -0.75$, $\phi^{\ast} = 90^{\circ}$, $v_{\text{cut}} = 0.166~\mathrm{GeV}^{2}$. Curves: $Q^{2} = 0.4105, 0.7085, 1.6980, 4.0855~\mathrm{GeV}^{2}$ (see legend). The horizontal line marks $\delta = 1$. All curves rise from $\delta \approx 0.61 - 0.64$ at threshold to a local maximum at $W \simeq 1.655~\mathrm{GeV}$ (peak $\delta \approx 1.05 - 1.25$, increasing with $Q^{2}$), then fall and flatten to $\delta \approx 0.86 - 0.90$ for $W \gtrsim 1.85~\mathrm{GeV}$.}
  \label{fig:delta_W_multiQ2}
\end{figure}

\begin{figure}[tbp]
  \centering
  \includegraphics[width=\linewidth]{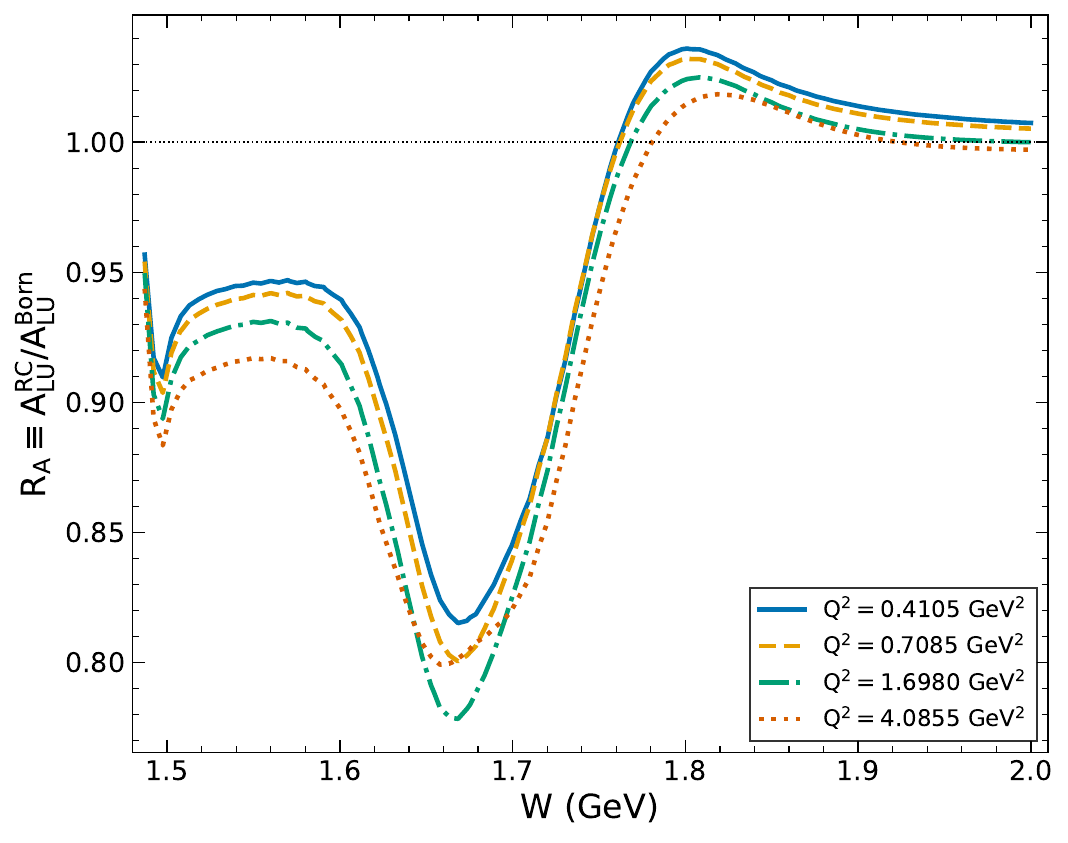}
  \caption{Asymmetry ratio $R_A \equiv A^{\rm RC}_{LU}/A^{\rm Born}_{LU}$ vs.\ $W$ at $\cos\theta^{\ast} = -0.75$, $\phi^{\ast} = 90^{\circ}$, $v_{\text{cut}} = 0.166~\mathrm{GeV}^{2}$. Curves: $Q^{2} = 0.4105$, $0.7085$, $1.6980$, $4.0855~\mathrm{GeV}^{2}$ (see legend). Near threshold: $R_A \approx 0.88 - 0.91$. Local minimum $R_A \approx 0.78 - 0.82$ at $W \simeq 1.66 - 1.68~\mathrm{GeV}$; local maximum $R_A \approx 1.02 - 1.04$ near $W \simeq 1.80 - 1.82~\mathrm{GeV}$; plateau $R_A \approx 1.00$ for $W \gtrsim 1.9~\mathrm{GeV}$.}
  \label{fig:RA_W_multiQ2}
\end{figure}

Figure~\ref{fig:delta_W_multiQ2} shows the cross-section RC factor $\delta$ versus $W$ at fixed $(\cos\theta^{\ast}, \phi^{\ast}) = (-0.75,\, 90^{\circ})$ for four values of $Q^{2}$. All four curves rise from $\delta \approx 0.61 - 0.64$ at the threshold to a local maximum at $W \simeq 1.655~\mathrm{GeV}$, where $\delta$ exceeds unity for all $Q^{2}$ values; the peak height increases from $\delta \approx 1.05$ at $Q^{2} = 0.4105~\mathrm{GeV}^{2}$ to $\delta \approx 1.25$ at $Q^{2} = 4.0855~\mathrm{GeV}^{2}$. The peak position at $W \approx 1.655$ GeV is the same for all four $Q^2$ values; it is pinned by the resonance content of the hadronic amplitude input rather than by the photon phase space. Beyond the peak, the curves fall through unity and flatten to a plateau of $\delta \approx 0.86 - 0.90$ for $W \gtrsim 1.85~\mathrm{GeV}$.
Figure~\ref{fig:RA_W_multiQ2} shows the asymmetry ratio $R_A$ at the same fixed angles. Near the threshold, the curves lie at $R_A \approx 0.88 - 0.91$, fall to a local minimum of $R_A \approx 0.78 - 0.82$ at $W \simeq 1.66 - 1.68~\mathrm{GeV}$, rise to a local maximum of $R_A \approx 1.02 - 1.04$ near $W \simeq 1.80 - 1.82~\mathrm{GeV}$, and level off near $R_A \approx 1.00$ for $W \gtrsim 1.9~\mathrm{GeV}$. The minimum is deepest at $Q^2 = 1.698~\mathrm{GeV}^2$ ($R_A \approx 0.779$) rather than at the highest $Q^2$ value ($R_A \approx 0.800$ at $Q^2 = 4.0855~\mathrm{GeV}^2$), so the suppression at the $R_A$ minimum is not monotonic across the $Q^2$ range shown. The spread across the four $Q^{2}$ curves is at most a few percent at any given $W$.
The two RC factors $\delta$ and $R_A$ are, in general, independent and coincide at a given kinematic point only by accident. Comparing Figs.~\ref{fig:delta_W_multiQ2} and~\ref{fig:RA_W_multiQ2}, $R_A$ reaches its minimum where $\delta$ peaks, and vice versa. This anti-correlation is consistent with the pion-channel behavior in Figs.~7--9 of Ref.~\cite{Afanasev:2002ee}, where the beam spin asymmetry RC changes sign between the $\Delta(1232)$ and $S_{11}(1535)$ resonance regions. The mechanism is the same here: the hard-bremsstrahlung integral enters the polarized and unpolarized cross-section pieces with different weights at kinematics where the hadronic amplitudes vary rapidly with $W$, so $R_A$ departs from unity wherever $\delta$ is farthest from unity.

The $W$-dependence of $\delta$ in the $\eta$ channel is more structured than the pion results in Figs.~3--8 of Ref.~\cite{Afanasev:2002ee}. This follows from the EtaMAID resonance content: $\eta$ production near threshold is dominated by the $S_{11}(1535)$ and $S_{11}(1650)$ resonances, whose contributions to the hadronic model input are concentrated in a narrow $W$ interval. Pion production in the same $W$ range involves several overlapping resonances, which smooth out the $W$-dependence of $\delta$. The sharper structure in the $\eta$ channel is therefore a feature of the hadronic physics, not an artifact of the RC calculation.

%===================================================================
\subsection{Angular dependence}\label{sec:results_angular}
\begin{figure}[tbp]
    \centering
    \includegraphics[width=\linewidth]{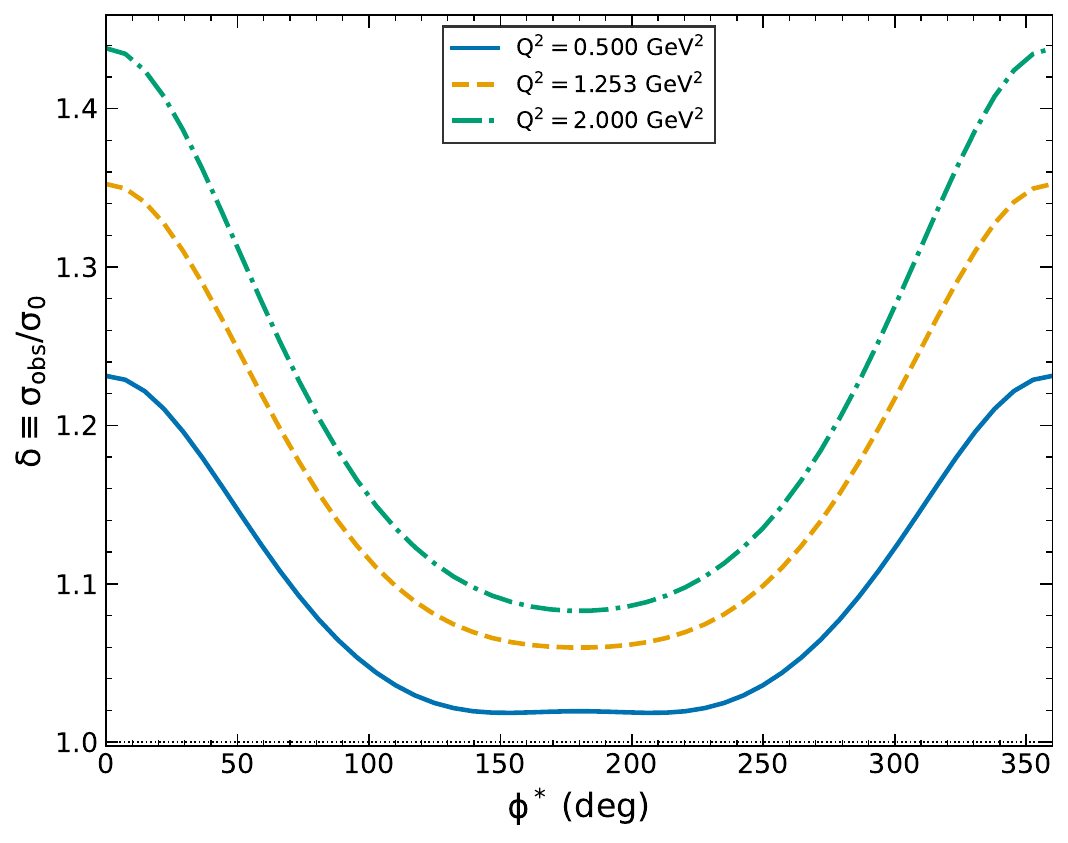}
    \caption{$\delta \equiv \sigma_{\rm obs}/\sigma_{0}$ vs.\ $\phi^{\ast}$ at $W = 1.660~\mathrm{GeV}$, $\cos\theta^{\ast} = -0.750$, $v_{\text{cut}} = 0.166~\mathrm{GeV}^{2}$, for $Q^{2} = 0.500, 1.253, 2.000~\mathrm{GeV}^{2}$ (see legend). Cosine-like modulation with $\delta > 1$ throughout; both amplitude and offset increase with $Q^{2}$. At $Q^{2} = 0.500~\mathrm{GeV}^{2}$, $\delta \in [1.02, 1.23]$; at $Q^{2} = 2.000~\mathrm{GeV}^{2}$, $\delta \in [1.08, 1.44]$. The horizontal dotted line marks $\delta = 1$.}
    \label{fig:delta_phi_main}
\end{figure}

\begin{figure}[tbp]
    \centering
    \includegraphics[width=\linewidth]{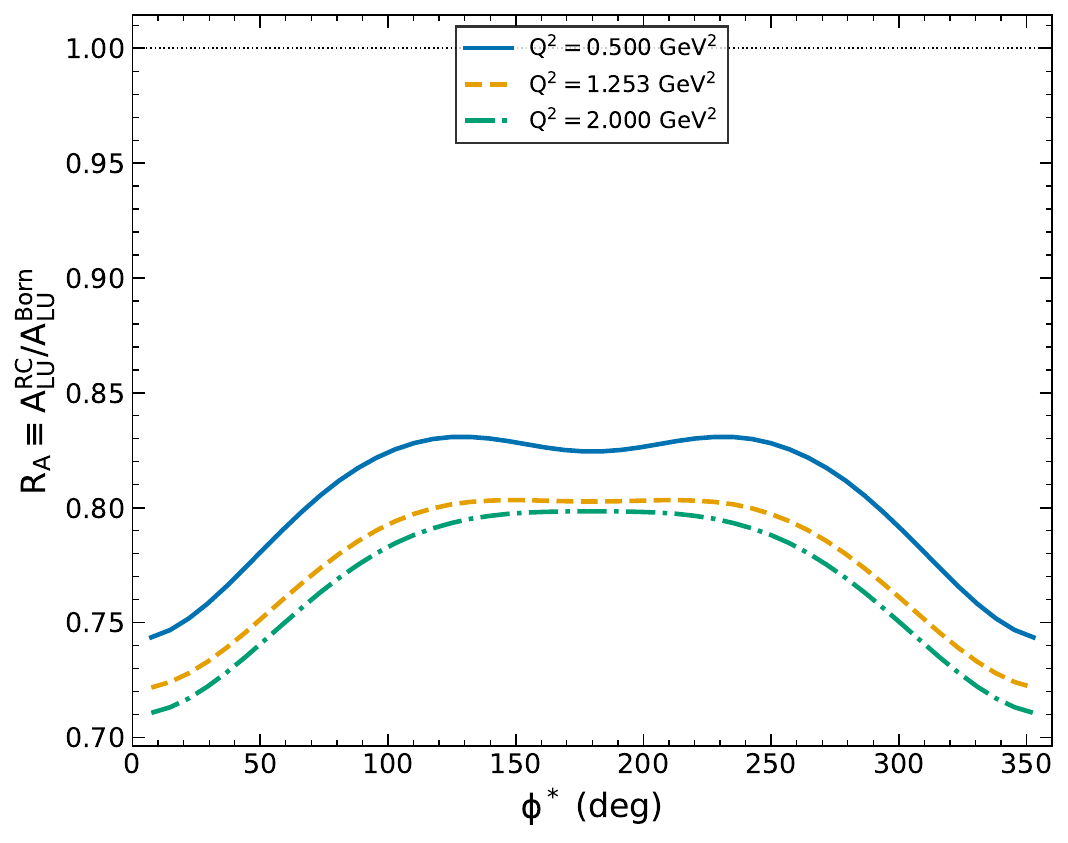}
    \caption{Asymmetry ratio $R_A \equiv A_{LU}^{\rm RC}/A_{LU}^{\rm Born}$ vs.\ $\phi^{\ast}$ at $W = 1.660~\mathrm{GeV}$, $\cos\theta^{\ast} = -0.750$, $v_{\text{cut}} = 0.166~\mathrm{GeV}^{2}$, for $Q^{2} = 0.500$, $1.253$, $2.000~\mathrm{GeV}^{2}$ (see legend). The dotted horizontal line marks $R_A = 1$ (no RC effect on the asymmetry). $R_A$ is undefined near $\phi^{\ast} = 0^{\circ}$ and $180^{\circ}$ where $A_{LU}^{\rm Born} \to 0$. All curves lie below unity; suppression deepens with $Q^{2}$: $R_A \in [0.74,\,0.83]$ at $Q^{2} = 0.500~\mathrm{GeV}^{2}$ and $R_A \in [0.71,\,0.80]$ at $Q^{2} = 2.000~\mathrm{GeV}^{2}$. The weak $\phi^{\ast}$ modulation, peaking near $\phi^{\ast} \approx 90^{\circ}$, indicates that the $\sin\phi^{\ast}$ shape of $A_{LU}$ is preserved under radiative corrections.}
    \label{fig:RA_phi_main}
\end{figure}

Figure~\ref{fig:delta_phi_main} shows $\delta$ as a function of $\phi^{\ast}$ at $W = 1.660~\mathrm{GeV}$, the $\delta(W)$ local maximum identified in Fig.~\ref{fig:delta_W_multiQ2}, for three values of $Q^{2}$ at fixed $\cos\theta^{\ast} = -0.750$. The correction follows a cosine-like modulation with $\delta > 1$ throughout; hard-photon emission enhances the observed cross section relative to the Born approximation at these kinematics. Both the amplitude and offset of the modulation increase with $Q^{2}$: at $Q^{2} = 0.500~\mathrm{GeV}^{2}$ the correction ranges from $\delta \approx 1.02$ at $\phi^{\ast} \simeq 180^{\circ}$ to $\delta \approx 1.23$ at $\phi^{\ast} \simeq 0^{\circ}$, while at $Q^{2} = 2.000~\mathrm{GeV}^{2}$ the range extends to $\delta \approx 1.08 - 1.44$.
Figure~\ref{fig:RA_phi_main} shows the asymmetry ratio $R_A \equiv A_{LU}^{\rm RC}/A_{LU}^{\rm Born}$ as a function of $\phi^{\ast}$ at the same kinematics as Fig.~\ref{fig:delta_phi_main}. $R_A$ is undefined near $\phi^{\ast} = 0^{\circ}$ and $180^{\circ}$ where $A_{LU}^{\rm Born} \to 0$. All three curves lie below unity throughout the accessible range, confirming that RC systematically suppresses the asymmetry at $W = 1.660~\mathrm{GeV}$. Each curve shows a weak $\phi^{\ast}$ modulation with a broad maximum near $\phi^{\ast} \approx 90^{\circ}$: $R_A$ ranges from ${\sim}\,0.74$--$0.83$ at $Q^{2} = 0.500~\mathrm{GeV}^{2}$ and from ${\sim}\,0.71$--$0.80$ at $Q^{2} = 2.000~\mathrm{GeV}^{2}$. The suppression deepens with $Q^{2}$, consistent with the larger $\delta$ at higher $Q^{2}$ in Fig.~\ref{fig:delta_phi_main}. The near-uniformity of $R_A$ across $\phi^{\ast}$ confirms that the $\sin\phi^{\ast}$ form of $A_{LU}$ is preserved under radiative corrections; the $\sin\phi^{\ast}$ moment extraction described in \hyperref[SM:S3]{SM~S3} is therefore unaffected by RC at this level.

%==============================================================
\subsection{Dependence on the inelasticity cut}\label{sec:results_vcut}
\begin{figure}[tbp]
    \centering
    \includegraphics[width=\linewidth]{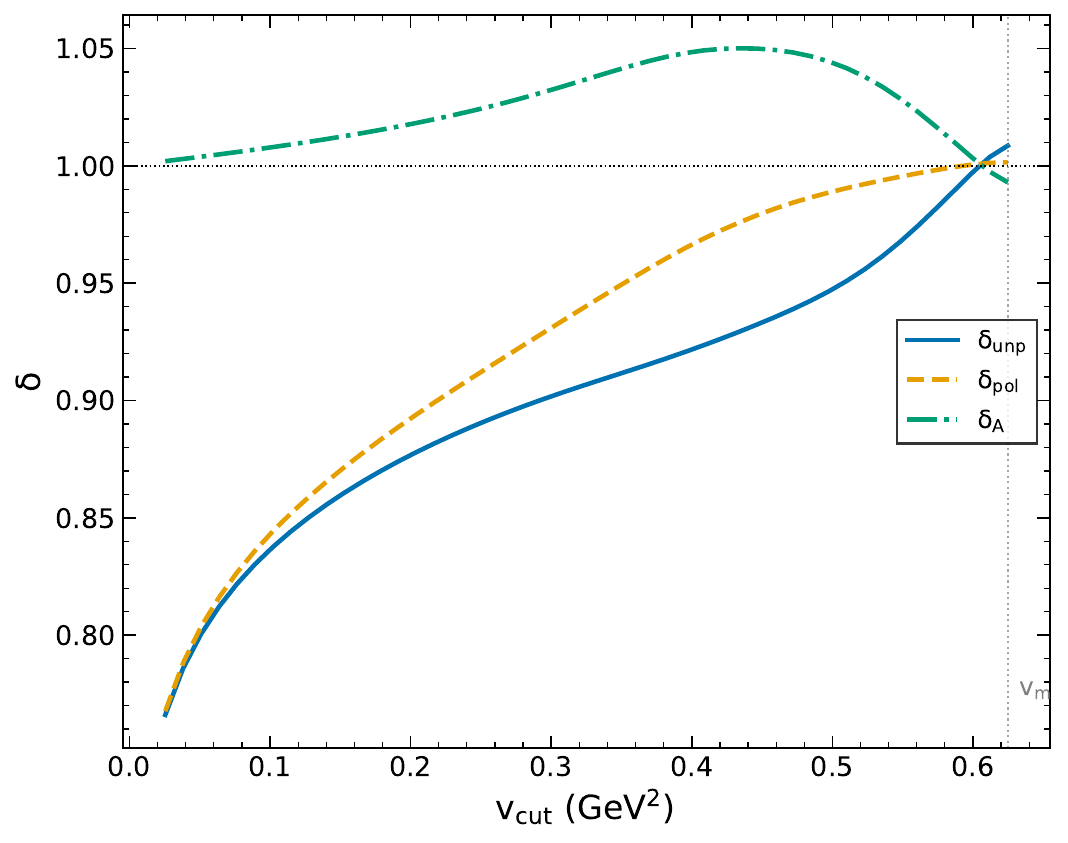}
    \caption{RC factors vs.\ inelasticity cut $v_{\text{cut}}$ for $\eta$ electroproduction at $W = 1.900~\mathrm{GeV}$, $Q^{2} = 0.4105~\mathrm{GeV}^{2}$, $\cos\theta^{\ast} = -0.750$, $\phi^{\ast} = 90^{\circ}$. $\delta_{\text{unp}}$: RC to the unpolarized cross section; $\delta_{\text{pol}}$: RC to the polarized cross section; $\delta_A = \delta_{\text{unp}}/\delta_{\text{pol}}$: RC to the beam polarization asymmetry. Kinematic upper limit $v_m \approx 0.625~\mathrm{GeV}^{2}$ (Eq.~(\ref{eq:vmax})). EtaMAID was used for structure functions.}
    \label{fig:vcut_scan}
\end{figure}

Figure~\ref{fig:vcut_scan} shows the dependence of the RC on the inelasticity cut $v_{\text{cut}}$. For $\eta$ electroproduction with a detected proton, the kinematic maximum of the inelasticity is $v_m = v_{\rm max}$ from Eq.~(\ref{eq:vmax}). At $W = 1.900~\mathrm{GeV}$, this gives $v_m \approx 0.625~\mathrm{GeV}^{2}$. The figure shows the RC factors for the unpolarized cross section ($\delta_{\text{unp}}$), the polarized cross section ($\delta_{\text{pol}}$), and the beam spin asymmetry ($\delta_A = \delta_{\text{unp}}/\delta_{\text{pol}}$) as functions of $v_{\text{cut}}$. For small $v_{\text{cut}}$, only soft bremsstrahlung contributes; the corrections to the polarized and unpolarized pieces are similar, and $\delta_A$ stays close to unity. As $v_{\text{cut}}$ increases, hard-photon emission enters, and the two corrections separate: $\delta_{\text{unp}}$ rises from ${\sim}\,0.77$ toward unity. In contrast, $\delta_{\text{pol}}$ rises more slowly, producing $\delta_A > 1$ in the mid-range before the three curves converge near $v_{\text{cut}} = v_m$. This pattern matches the pion-channel result in Fig.~9 of Ref.~\cite{Afanasev:2002ee}.

Extended kinematic scans are collected in \hyperref[SM:S2]{SM~S2}. Two results 
warrant attention: the $\cos\theta^{\ast}$ scan (SM~S2.1) shows that $\delta$ reaches 
${\sim}\,1.76$ at forward angles ($\cos\theta^{\ast} = +0.9$) at the $\delta$-peak 
kinematics, while $R_A$ falls to ${\sim}\,0.67$ at the same point. The two effects 
are the largest deviations from unity in the full grid. The $Q^2$ scan (SM~S2.2) shows 
a localized non-monotonic feature near $Q^2 \approx 0.55$--$0.70$~GeV$^2$ that appears 
coherently across all four kinematic combinations examined and is attributed to 
EtaMAID-2023 multipole structure. The full RC grid is available interactively at 
\url{https://izzyillari.github.io/exclurad/} (\hyperref[SM:S4]{SM~S4}).

%=================================================================
%-------------------- Conclusion --------------------------------
%================================================================
\section{Conclusion}\label{sec:conclusion}
We extended EXCLURAD for exact $\mathcal{O}(\alpha)$ QED radiative corrections from exclusive pion electroproduction to the $\eta$ channel $e\,p \to e^\prime\,p\,\eta (\gamma)$. The modifications were restricted to kinematics ($m_\pi \to m_\eta$), numerical precision, and the model-input interface; the underlying Bardin-Shumeiko formalism and all analytic expressions remain unchanged. An explicit leading-logarithm formula for the $\eta$ channel, absent from Ref.~\cite{Afanasev:2002ee}, is derived in Sec.~\ref{sec:ll}. Structure functions were supplied by EtaMAID-2023 multipole tables covering $W = 1.486$--$2.0$~GeV and $Q^{2} = 0$--$5$~GeV$^{2}$; the model is most reliable for $W \lesssim 2$~GeV, where the $s$-channel resonance description applies.

The cross-section RC factor $\delta$ varies by up to ${\sim}\,30\%$ across the resonance region, with a local maximum near $W \simeq 1.66$~GeV whose height increases with $Q^{2}$, and a sub-unity plateau of $\delta \simeq 0.86$--$0.92$ at higher $W$. The asymmetry ratio $R_A$ is anti-correlated with $\delta$, consistent with the pion-channel behavior of Ref.~\cite{Afanasev:2002ee} and arising from the different RC weights on the polarized and unpolarized pieces in the hard-bremsstrahlung integral. The $W$-dependence of $\delta$ is more structured than in the pion channel because the $S_{11}(1535)$ and $S_{11}(1650)$ resonances dominate $\eta$ production over a narrow $W$ interval, unlike the overlapping resonances in the pion case. The $\sin\phi^{\ast}$ form of $A_{LU}$ is preserved under RC, and the $v_{\text{cut}}$ dependence follows the pattern established in the pion channel. The nontrivial angular dependence of $\delta$ in both $\cos\theta^{\ast}$ and $\phi^{\ast}$ means that radiative corrections must be applied differentially across the full kinematic grid rather than as a single multiplicative factor; this requirement is more pronounced here than in the pion channel because of the compressed phase space between the $\eta$ threshold and the dominant resonance.

Model dependence in $\delta$ is expected to be small in the resonance-dominated region, by analogy with the pion-channel study in Ref.~\cite{Afanasev:2002ee}, where three hadronic models agreed to within a few percent near the $\Delta(1232)$ peak. This estimate is a lower bound: the pion-channel spread reached ${\sim}\,5\%$ even near the dominant resonance, and the $\eta$ channel carries additional uncertainty given that EtaMAID-2023 predictions for $A_{LU}$ differ in sign from measured values~\cite{Illari:2024lvw}. A quantitative model-comparison study, requiring EXCLURAD-compatible multipole tables from the JBW~\cite{Mai2021, Mai2022, Mai2023, Wang2024} or Bonn-Gatchina frameworks, is left for future work.

The $\eta$ and $\eta^\prime$ channels share similar production dynamics. EtaMAID-2023 multipole tables for the $\eta^\prime p$, $\eta^\prime n$, and $\eta n$ channels are available from the same framework~\cite{Tiator2018}, so extending EXCLURAD to $\eta^\prime$ electroproduction requires only the mass substitution and the generation of the corresponding lookup tables.

%===============================================================
%------------- Data and Code Availability ---------------------
%==============================================================
\section{Data and Code Availability}\label{sec:data_code}
The modified EXCLURAD source, EtaMAID-derived lookup tables, run scripts, and plotting code used in this work are available at \url{https://github.com/IzzyIllari/exclurad} and archived at DOI:~\href{https://doi.org/10.5281/zenodo.18970108}{10.5281/zenodo.18970108} (v1.0.0). An interactive browser-based tool for exploring the radiative-correction results across the full $(W, Q^{2}, \cos\theta^{\ast}, \phi^{\ast})$ grid is deployed at \url{https://izzyillari.github.io/exclurad/} and described in \hyperref[SM:S4]{SM~S4}. The CERNLIB-free distribution of EXCLURAD is maintained at \url{https://github.com/JeffersonLab/exclurad}; legacy materials and earlier documentation are available at \url{https://www.jlab.org/RC/}. File checksums and command-line recipes for reproducing all figures are provided in \hyperref[SM:S4]{SM~S4}.

%==========================================================
%--------------- Acknowledgments --------------------------
%===========================================================
\section{Acknowledgments}

This work was supported in part by the U.S. Department of Energy, Office of Science, Office of Nuclear Physics, under Award No.\ DE--SC0016583. A.~A. acknowledges support from the U.S. National Science Foundation under grants PHY--2111063 and PHY--2514669.

%=============================================================
%---------------------- REFERENCES -------------------------
%==============================================================
\bibliographystyle{apsrev4-1}
\bibliography{references}

%========================= SM ================================
\clearpage
\onecolumngrid
\appendix
\section*{Supplementary Material}
% ===============================================================
%                   SUPPLEMENTARY MATERIAL
%              (publication version)
% ===============================================================
% This file is \input'd from the main document after \references.
% Sections are numbered S1--S4.
% ===============================================================

% ================================================================
% SM S1 — Extended vcut scans
% ================================================================
\subsection*{S1. Extended \texorpdfstring{$v_{\mathrm{cut}}$}{vcut} variations}\label{SM:S1}
The main-text figure (Figure ~\ref{fig:vcut_scan}) shows the dependence of the RC factors on the inelasticity cut $v_{\mathrm{cut}}$ at a single kinematic point: $W = 1.900$~GeV, $Q^{2} = 0.4105~\mathrm{GeV}^{2}$, $\cos\theta^{\ast} = -0.750$, $\phi^{\ast} = 90^{\circ}$. This section extends that scan to a range of $W$ and $Q^{2}$ values at the same fixed angles.

In all figures, the three RC factors are defined as in the main text: $\delta_{\mathrm{unp}} = \sigma_{\mathrm{obs}}/\sigma_0$ for the unpolarized cross section, $\delta_{\mathrm{pol}}$ for the helicity-difference cross section, and $\delta_A = \delta_{\mathrm{unp}}/\delta_{\mathrm{pol}}$ for the beam polarization asymmetry. All three converge to their respective physical values at the kinematic upper limit $v_m(W) = (W - M_p)^2 - m_\eta^2$ (dotted vertical line), which depends only on $W$ and is independent of $Q^2$.

Each panel uses a shared $x$-axis extending from $v_{\mathrm{cut}} = 0.02~\mathrm{GeV}^2$ to $v_m(W_{\mathrm{max}}) \times 1.06$, where $W_{\mathrm{max}}$ is the largest $W$ among all panels in that figure. Panels with smaller $W$ therefore show data that end well before the right edge of the frame; the dotted vertical line marks the $v_m$ appropriate for each panel. The larger $v_m$ in the $W$-variation panels is due to the wider bremsstrahlung phase space at higher $W$.

%--------------------------------------------------------------
\subsubsection*{\texorpdfstring{$Q^2$ dependence at fixed $W$}{Q2 dependence at fixed W}}
\begin{figure*}[!htbp]
    \centering
    \includegraphics[width=\textwidth]{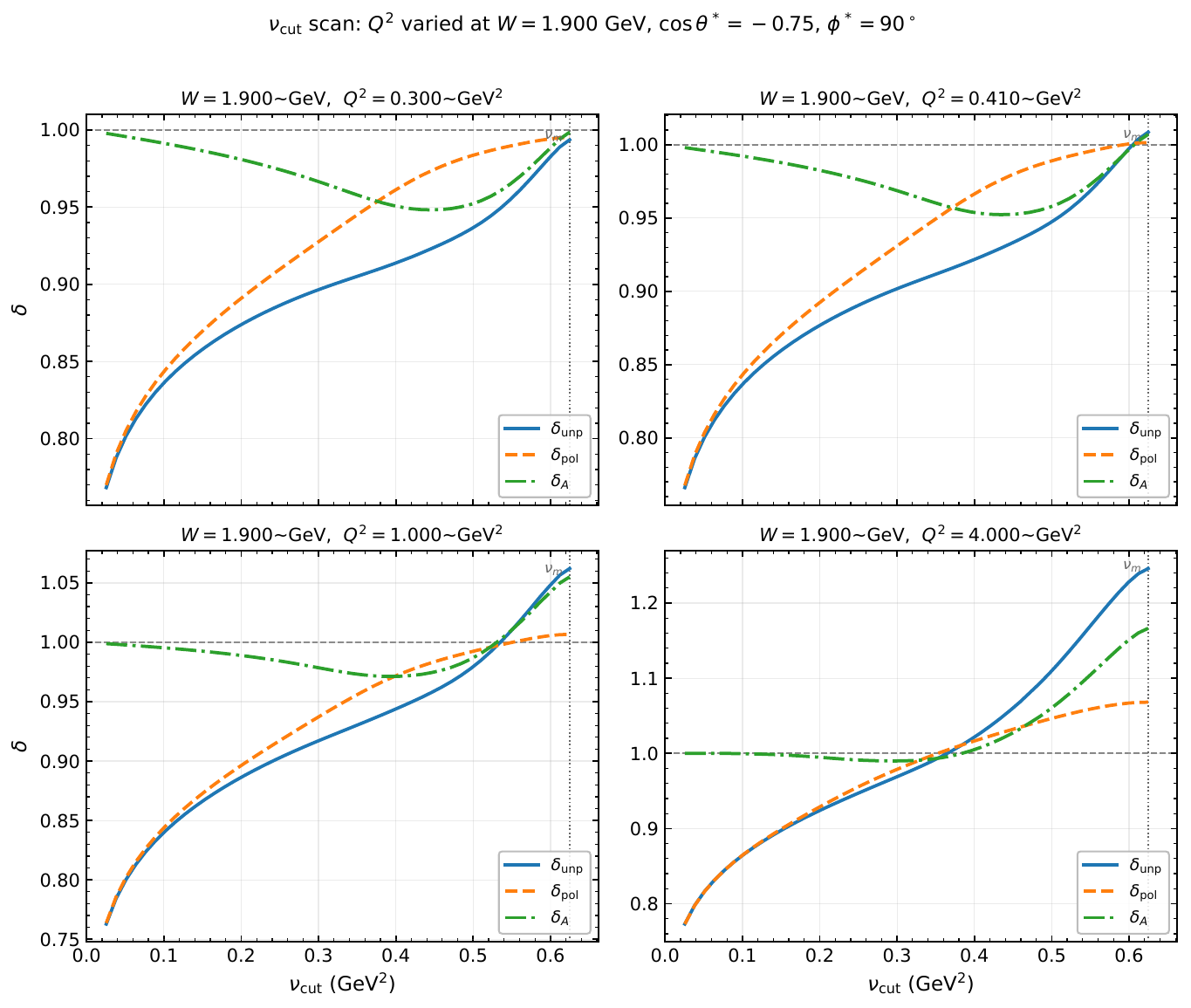}
    \caption{RC factors $\delta_{\mathrm{unp}}$ (blue solid), $\delta_{\mathrm{pol}}$ (orange dashed), and $\delta_A = \delta_{\mathrm{unp}}/\delta_{\mathrm{pol}}$ (green dash-dot) vs.\ $v_{\mathrm{cut}}$ at $W = 1.900$~GeV, $\cos\theta^{\ast} = -0.750$, $\phi^{\ast} = 90^{\circ}$, for four values of $Q^2$ (panels). The dashed horizontal line marks $\delta = 1$; the dotted vertical line marks $v_m \approx 0.625~\mathrm{GeV}^{2}$. The shared $x$-axis extends to $v_m \times 1.06$. EtaMAID was used for structure functions.}
    \label{fig:s4_Q2_scan}
\end{figure*}

Figure~\ref{fig:s4_Q2_scan} shows the $v_{\mathrm{cut}}$ scan at $W = 1.900$~GeV for $Q^{2} \in \{0.300,\, 0.410,\, 1.000,\, 4.000\}~\mathrm{GeV}^{2}$. All four panels share $v_m \approx 0.625~\mathrm{GeV}^{2}$. The qualitative pattern established in the main text is stable: $\delta_{\mathrm{unp}}$ and $\delta_{\mathrm{pol}}$ both rise from $\sim 0.77$ at small $v_{\mathrm{cut}}$ toward their $v_m$ values. At the same time, $\delta_A$ is non-monotonic, peaking above unity at intermediate $v_{\mathrm{cut}}$ before returning to $\sim 1$ as the full radiative tail is included. The $Q^2$ dependence enters through the hadronic model (EtaMAID multipoles) in the hard-photon integrand. At low $Q^2$ ($0.300$ and $0.410~\mathrm{GeV}^{2}$) the separation between $\delta_{\mathrm{unp}}$ and $\delta_{\mathrm{pol}}$ is modest, and $\delta_A$ remains within a few percent of unity across most of the $v_{\mathrm{cut}}$ range. At high $Q^2$ ($1.000$ and $4.000~\mathrm{GeV}^{2}$) the separation grows substantially: at $Q^2 = 4.000~\mathrm{GeV}^{2}$, $\delta_{\mathrm{unp}}$ overshoots unity and reaches $\sim 1.25$ at $v_m$, while $\delta_{\mathrm{pol}} \approx 1.07$ gives $\delta_A \approx 1.17$. The RC effect on the asymmetry, therefore, grows with $Q^2$ at fixed $W$.

%-----------------------------------------------------------------
\subsubsection*{\texorpdfstring{$W$ dependence at fixed $Q^2$}{W dependence at fixed Q2}}
\begin{figure*}[!htbp]
    \centering
    \includegraphics[width=\textwidth]{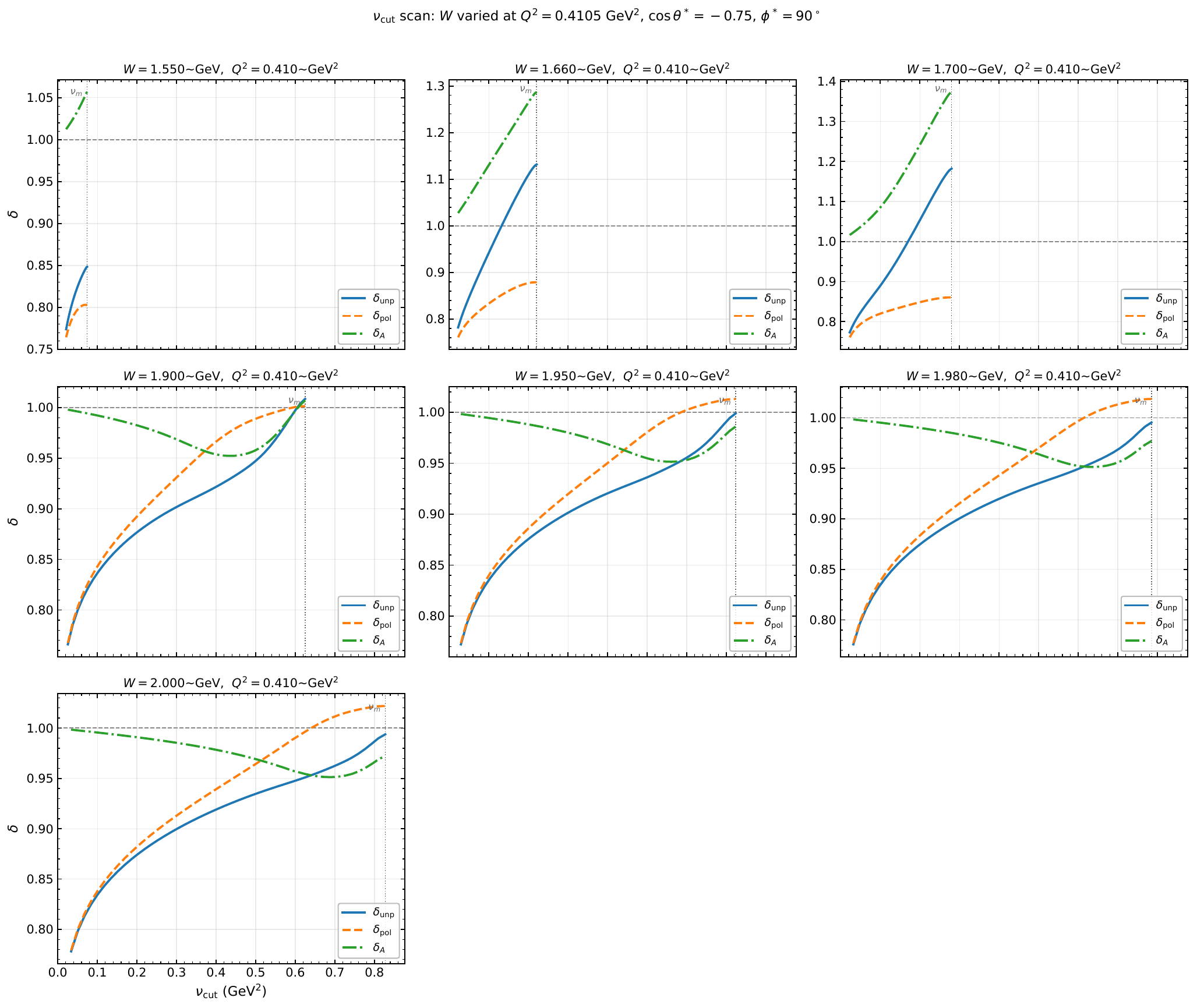}
    \caption{Same as Figure ~\ref{fig:s4_Q2_scan} but at $Q^2 = 0.4105~\mathrm{GeV}^{2}$ for seven values of $W$ (panels). The dotted vertical line in each panel marks the corresponding $v_m(W) = (W-M_p)^2 - m_\eta^2$, which ranges from $0.074~\mathrm{GeV}^{2}$ at $W = 1.550$~GeV to $0.827~\mathrm{GeV}^{2}$ at $W = 2.000$~GeV. The shared $x$-axis extends to $v_m(W=2.000) \times 1.06 \approx 0.877~\mathrm{GeV}^{2}$; panels with smaller $W$ show data ending before the right edge of the frame.}
    \label{fig:s4_W_scan}
\end{figure*}

Figure~\ref{fig:s4_W_scan} shows the $v_{\mathrm{cut}}$ scan at $Q^{2} = 0.4105~\mathrm{GeV}^{2}$ for $W \in \{1.550,\, 1.660,\, 1.700,\, 1.900,\, 1.950,\, 1.980,\, 2.000\}$~GeV. Near threshold ($W = 1.550~\mathrm{GeV}$), the kinematic window is small ($v_m \approx 0.074~\mathrm{GeV}^{2}$), so the scan covers only a narrow range, and the corrections are compact. At $W = 1.660~\mathrm{GeV}$, which corresponds to the region where the main-text $\delta(W)$ scan shows its local maximum (cf.\ Figure ~\ref{fig:delta_W_multiQ2}), $\delta_A$ reaches $\sim 1.3$ at $v_m$, the largest excursion across all panels. This follows from the rapid variation of the hadronic model input near the $S_{11}(1535)$ and $S_{11}(1650)$ resonances. From $W = 1.900~\mathrm{GeV}$ onward, the shape converges to the pattern shown in the main text: a concave rise of $\delta_{\mathrm{unp}}$ and $\delta_{\mathrm{pol}}$ from $\sim 0.77$, a broad arch of $\delta_A$ above unity in the mid-range, and convergence near $v_m$. The arch of $\delta_A$ becomes slightly more pronounced at higher $W$, where the bremsstrahlung phase space is larger.

\clearpage

% ================================================================
% SM S2 -- Additional delta, A_LU, and R_A scans
% ================================================================
\subsection*{S2. Additional \texorpdfstring{$\delta$}{delta},
\texorpdfstring{$A_{LU}$}{ALU}, and \texorpdfstring{$R_A$}{RA}
kinematic dependence}\label{SM:S2}
This section collects supplemental kinematic scans in four parts.
\hyperref[SM:S2.1]{S2.1} presents $\delta$, $A_{LU}$, and
$R_A \equiv A_{LU}^{\rm RC}/A_{LU}^{\rm Born}$ as functions of
$\cos\theta^{\ast}$ at two values of $W$: the $\delta$ peak at
$W = 1.660~\mathrm{GeV}$ and the post-peak plateau at
$W = 1.900~\mathrm{GeV}$, each for $Q^{2} = 0.500$, $1.253$, and
$2.000~\mathrm{GeV}^{2}$.
\hyperref[SM:S2.2]{S2.2} presents the same three quantities as
functions of $Q^{2}$ at $\phi^{\ast} = 90^{\circ}$ for four
$(W, \cos\theta^{\ast})$ kinematic combinations.
\hyperref[SM:S2.3]{S2.3} shows $\delta$ and $A_{LU}$ versus
$\phi^{\ast}$ at $W = 1.660~\mathrm{GeV}$ for both a $Q^{2}$ scan
and a $W$ scan, extending the main-text Figs.~\ref{fig:delta_phi_main}
and~\ref{fig:RA_phi_main} to a fuller kinematic survey.
\hyperref[SM:S2.4]{S2.4} shows $R_A$ versus $\phi^{\ast}$ for the
same two scan sets, complementing the $W$-dependent $R_A$ of
Figure ~\ref{fig:RA_W_multiQ2}.
All figures use $v_{\mathrm{cut}} = 0.166~\mathrm{GeV}^{2}$ and
$\phi^{\ast} = 90^{\circ}$ (S2.1 and S2.2) or scan over $\phi^{\ast}$
(S2.3 and S2.4).

% ------------------------------------------------------------------
\subsubsection*{\texorpdfstring{S2.1\quad Angular scans: $\delta$, $A_{LU}$, and $R_A$ versus $\cos\theta^{\ast}$}{S2.1 Angular scans}}\label{SM:S2.1}
\begin{figure*}[tbp]
  \centering
  \includegraphics[width=0.8\textwidth]{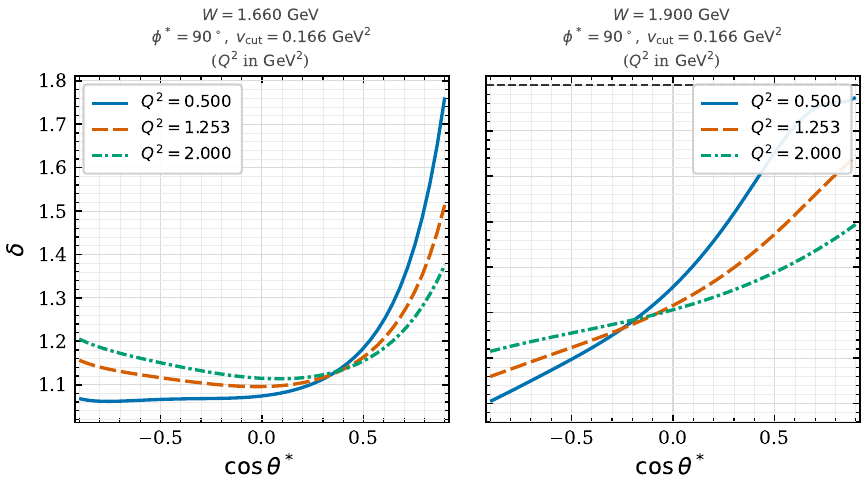}
  \caption{Cross-section RC factor $\delta$ vs.\ $\cos\theta^{\ast}$ at $\phi^{\ast} = 90^{\circ}$, $v_{\mathrm{cut}} = 0.166~\mathrm{GeV}^{2}$, for $Q^{2} = 0.500$, $1.253$, $2.000~\mathrm{GeV}^{2}$ (see legend). Dashed horizontal line: $\delta = 1$. Left: $W = 1.660~\mathrm{GeV}$ ($\delta$-peak kinematics). $\delta > 1$ throughout; at backward angles the curves are ordered $Q^{2} = 0.500 < 1.253 < 2.000$, spanning $\delta \approx 1.07 - 1.21$, and the ordering reverses near $\cos\theta^{\ast} \approx +0.4$. At $\cos\theta^{\ast} = +0.9$, $Q^{2} = 0.500~\mathrm{GeV}^{2}$ reaches $\delta \approx 1.76$ while $Q^{2} = 2.000~\mathrm{GeV}^{2}$ reaches $\delta \approx 1.38$. Right: $W = 1.900~\mathrm{GeV}$ (post-peak plateau). $\delta < 1$ throughout; the same inversion of $Q^{2}$ ordering occurs near $\cos\theta^{\ast} \approx +0.4$, but the total angular variation is modest ($\delta \approx 0.86 - 1.00$).}
  \label{fig:s21_delta_costh}
\end{figure*}

\begin{figure*}[tbp]
  \centering
  \includegraphics[width=0.8\textwidth]{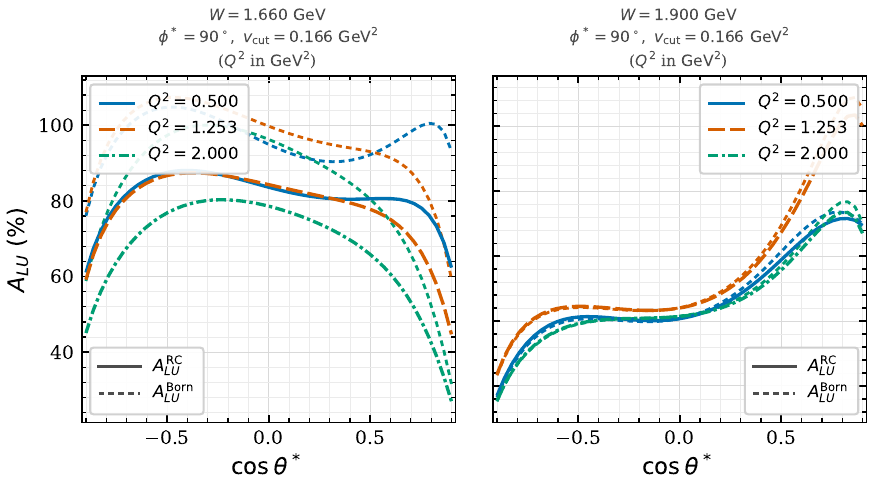}
  \caption{Beam-spin asymmetry $A_{LU}$ vs.\ $\cos\theta^{\ast}$ at $\phi^{\ast} = 90^{\circ}$, $v_{\mathrm{cut}} = 0.166~\mathrm{GeV}^{2}$, for $Q^{2} = 0.500$, $1.253$, $2.000~\mathrm{GeV}^{2}$ (see legend). Solid curves: $A_{LU}^{\rm RC}$; dashed curves: $A_{LU}^{\rm Born}$ (same color per $Q^{2}$). Left: $W = 1.660~\mathrm{GeV}$. Both RC and Born asymmetries peak near $\cos\theta^{\ast} \approx -0.4$; $A_{LU}^{\rm RC}$ lies below $A_{LU}^{\rm Born}$ at all angles, with the suppression growing at forward angles where $A_{LU}^{\rm Born}$ rises sharply for $Q^{2} = 1.253$ and $2.000~\mathrm{GeV}^{2}$ while their RC counterparts are more strongly suppressed. Right: $W = 1.900~\mathrm{GeV}$. The RC and Born curves are nearly coincident for each $Q^{2}$; both show a local minimum near $\cos\theta^{\ast} \approx 0$ and rise at backward and forward extremes. $Q^{2} = 1.253~\mathrm{GeV}^{2}$ produces the largest asymmetry at backward and central angles.}
  \label{fig:s21_alu_costh}
\end{figure*}

\begin{figure*}[tbp]
  \centering
  \includegraphics[width=0.8\textwidth]{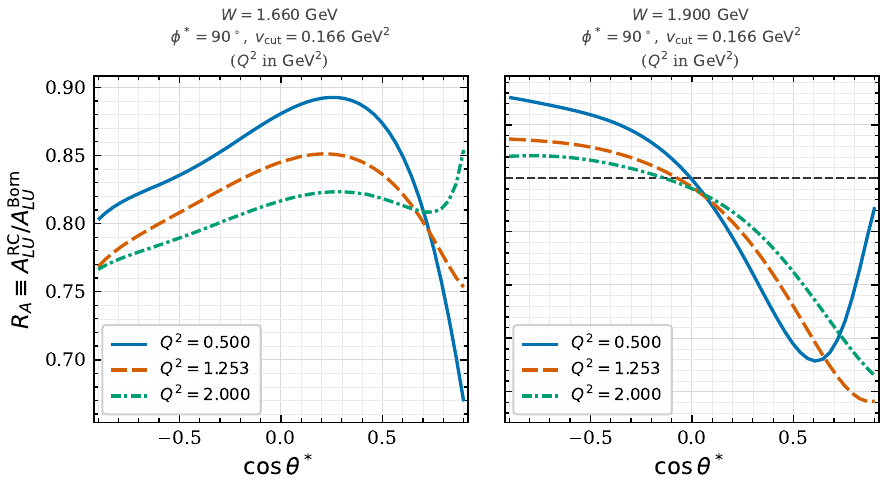}
  \caption{Asymmetry ratio $R_A \equiv A_{LU}^{\rm RC}/A_{LU}^{\rm Born}$ vs.\ $\cos\theta^{\ast}$ at $\phi^{\ast} = 90^{\circ}$, $v_{\mathrm{cut}} = 0.166~\mathrm{GeV}^{2}$, for $Q^{2} = 0.500$, $1.253$, $2.000~\mathrm{GeV}^{2}$ (see legend). Dashed horizontal line: $R_A = 1$ (no RC effect on asymmetry). Left: $W = 1.660~\mathrm{GeV}$. $R_A < 1$ throughout; each curve peaks near $\cos\theta^{\ast} \approx +0.1$--$+0.3$ and drops at very forward angles. $Q^{2} = 0.500~\mathrm{GeV}^{2}$ has the highest peak ($R_A \approx 0.89$) and the steepest forward drop ($R_A \approx 0.67$ at $\cos\theta^{\ast} = +0.9$); higher $Q^{2}$ gives deeper but more uniform suppression. The forward-angle drop reflects the same kinematic near-zero of $A_{LU}^{\rm Born}$ that produces the $R_A$ divergence in the $Q^{2}$ scan at fixed $\cos\theta^{\ast} = +0.750$ (\hyperref[SM:S2.2]{SM~S2.2}). The $R_A = 1$ reference line lies above the plotted range. Right: $W = 1.900~\mathrm{GeV}$. $R_A$ is close to 1 throughout with weak angular dependence; the reference line is visible near the top of the panel, and all curves remain within a few percent of it.}
  \label{fig:s21_ra_costh}
\end{figure*}

% ------------------------------------------------------------------
Figures~\ref{fig:s21_delta_costh}--\ref{fig:s21_ra_costh} show $\delta$, $A_{LU}$, and $R_A$ as functions of $\cos\theta^{\ast}$ at fixed $\phi^{\ast} = 90^{\circ}$, comparing results at the $\delta$ peak ($W = 1.660~\mathrm{GeV}$, left panels) and in the post-peak plateau ($W = 1.900~\mathrm{GeV}$, right panels). At $W = 1.660~\mathrm{GeV}$, $\delta$ exceeds unity across the full angular range and rises steeply toward forward angles (Figure ~\ref{fig:s21_delta_costh}, left). At backward angles ($\cos\theta^{\ast} \approx -0.9$) the three $Q^{2}$ curves are ordered $\delta(Q^{2} = 0.500) < \delta(Q^{2} = 1.253) < \delta(Q^{2} = 2.000)$, spanning roughly $1.07$--$1.21$. The ordering reverses near $\cos\theta^{\ast} \approx +0.4$, after which $Q^{2} = 0.500$ rises most steeply and reaches $\delta \approx 1.76$ at $\cos\theta^{\ast} = +0.9$, compared to $\delta \approx 1.38$ for $Q^{2} = 2.000$. At $W = 1.900~\mathrm{GeV}$ (Figure ~\ref{fig:s21_delta_costh}, right), $\delta < 1$ throughout; the same inversion of $Q^{2}$ ordering occurs at a similar crossover angle, but the total angular variation is modest, spanning roughly $0.86$--$1.00$ across the full $\cos\theta^{\ast}$ range.

The beam-spin asymmetry at $W = 1.660~\mathrm{GeV}$ (Figure ~\ref{fig:s21_alu_costh}, left) shows a broad maximum near $\cos\theta^{\ast} \approx -0.4$ for all three $Q^{2}$ values. The RC and Born curves are well separated: $A_{LU}^{\rm RC}$ lies below $A_{LU}^{\rm Born}$ across the full angular range, with the gap growing at forward angles where $A_{LU}^{\rm Born}$ diverges upward for $Q^{2} = 1.253$ and $2.000~\mathrm{GeV}^{2}$ while the corresponding RC curves are more strongly suppressed. At $W = 1.900~\mathrm{GeV}$ (Figure ~\ref{fig:s21_alu_costh}, right), $A_{LU}^{\rm RC}$ and $A_{LU}^{\rm Born}$ are nearly coincident for each $Q^{2}$, consistent with $R_A \approx 1$ in the plateau region. The angular shape shifts: both curves show a local minimum near $\cos\theta^{\ast} \approx 0$ and rise at both backward and forward extremes, with $Q^{2} = 1.253~\mathrm{GeV}^{2}$ producing the largest asymmetry at backward and central angles.

The ratio $R_A$ at $W = 1.660~\mathrm{GeV}$ (Figure ~\ref{fig:s21_ra_costh}, left) lies below 1 throughout, confirming RC suppression of the asymmetry at all angles. Each $Q^{2}$ curve reaches its maximum in the range $\cos\theta^{\ast} \approx +0.1$--$+0.3$ and then drops sharply at very forward angles: $Q^{2} = 0.500~\mathrm{GeV}^{2}$ peaks at $R_A \approx 0.89$ before falling to $R_A \approx 0.67$ at $\cos\theta^{\ast} = +0.9$, while higher $Q^{2}$ values show deeper but more uniform suppression across the angular range. This forward-angle suppression of $R_A$ occurs simultaneously with the large forward-angle enhancement of $\delta$: at $\cos\theta^{\ast} = +0.9$ and $W = 1.660~\mathrm{GeV}$, RC inflates the observed cross section by up to ${\sim}\,76\%$ (\textit{i.e.}, $\delta \approx 1.76$) while suppressing the asymmetry by up to ${\sim}\,33\%$ (\textit{i.e.}, $R_A \approx 0.67$). These two effects arise from different aspects of the QED correction and must be accounted for independently when applying RC to experimental observables. The sharp forward-angle drop in $R_A$ visible here also anticipates the behavior discussed in \hyperref[SM:S2.2]{SM~S2.2}: when $Q^{2}$ is scanned at fixed $\cos\theta^{\ast} = +0.750$, this same kinematic region is probed continuously, and $A_{LU}^{\rm Born}$ passes through a near-zero at large $Q^{2}$, causing $R_A$ to diverge. The $\cos\theta^{\ast}$ scan thus provides a cross-sectional view of the same near-zero that appears as a high-$Q^{2}$ divergence in the $Q^{2}$ scan of \hyperref[SM:S2.2]{SM~S2.2}.

At $W = 1.900~\mathrm{GeV}$ (Figure ~\ref{fig:s21_ra_costh}, right), $R_A$ is close to 1 across the full angular range, with weak angular dependence; the $R_A = 1$ reference line is visible near the top of the panel, and all curves remain within a few percent of it.

\clearpage

% ------------------------------------------------------------------
\subsubsection*{\texorpdfstring{S2.2\quad $Q^{2}$ scans: $\delta$, $A_{LU}$, and $R_A$ versus $Q^{2}$}{S2.2 Q2 scans}}\label{SM:S2.2}
\begin{figure*}[tbp]
  \centering
  \includegraphics[width=0.8\textwidth]{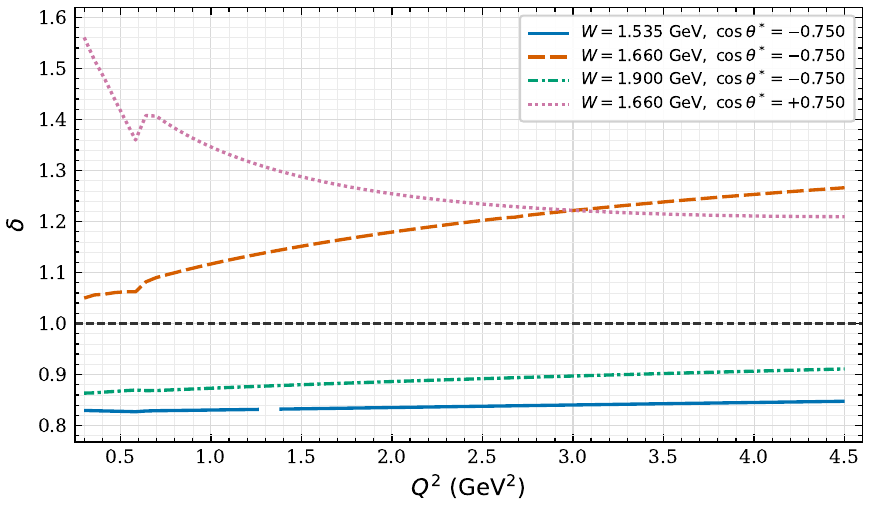}
  \caption{Cross-section RC factor $\delta$ vs.\ $Q^{2}$ at $\phi^{\ast} = 90^{\circ}$, $v_{\mathrm{cut}} = 0.166~\mathrm{GeV}^{2}$, for four kinematic combinations (see legend). Dashed horizontal line: $\delta = 1$. All four curves exhibit a localized non-monotonic feature near $Q^{2} \approx 0.55$--$0.70~\mathrm{GeV}^{2}$ that persists at 75-point resolution and is attributed to EtaMAID-2023 multipole structure (see text). The N(1535) ($W = 1.535~\mathrm{GeV}$) and post-peak plateau ($W = 1.900~\mathrm{GeV}$) curves have $\delta < 1$ throughout and rise slowly with $Q^{2}$. The $\delta$-peak backward curve ($W = 1.660~\mathrm{GeV}$, $\cos\theta^{\ast} = -0.750$) has $\delta > 1$ and increases monotonically from $1.050$ to $1.266$. The $\delta$-peak forward curve ($W = 1.660~\mathrm{GeV}$, $\cos\theta^{\ast} = +0.750$) starts at $\delta \approx 1.56$ and decreases; the two $W = 1.660~\mathrm{GeV}$ curves cross near $Q^{2} \approx 2.5~\mathrm{GeV}^{2}$. Small single-point gaps in three curves are EtaMAID-2023 interpolation nodes where the spline returns a non-finite value; each spans one grid step (${\sim}\,0.057~\mathrm{GeV}^{2}$) and does not affect the surrounding data.}
  \label{fig:s22_delta_q2}
\end{figure*}

\begin{figure*}[tbp]
  \centering
  \includegraphics[width=0.8\textwidth]{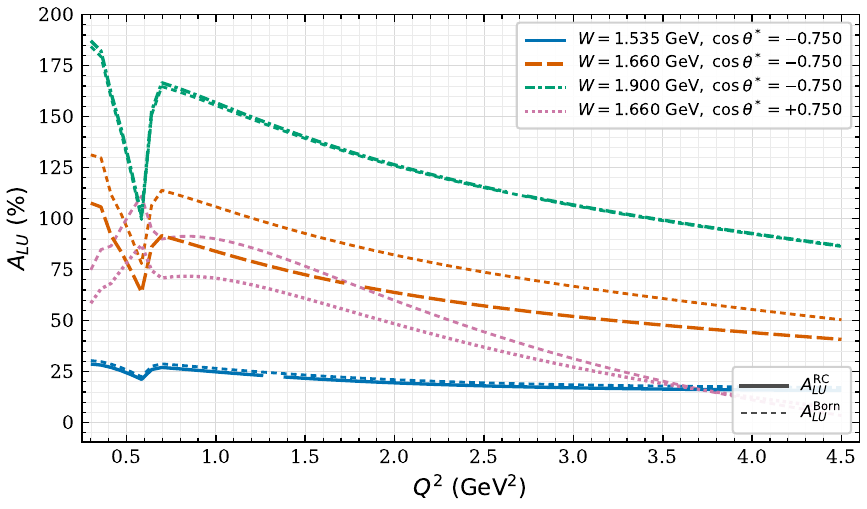}
  \caption{Beam-spin asymmetry $A_{LU}$ vs.\ $Q^{2}$ at $\phi^{\ast} = 90^{\circ}$, $v_{\mathrm{cut}} = 0.166~\mathrm{GeV}^{2}$, for four kinematic combinations (see legend). Solid curves: $A_{LU}^{\rm RC}$; short-dashed curves of the same color: $A_{LU}^{\rm Born}$. At $W = 1.535~\mathrm{GeV}$, both curves are small and nearly coincident. At $W = 1.900~\mathrm{GeV}$, the asymmetry exceeds $100\%$ at low $Q^{2}$, falls sharply through the bump region, and decreases to ${\sim}\,85\%$ at $Q^{2} = 4.50~\mathrm{GeV}^{2}$; RC and Born nearly coincide throughout. At both $W = 1.660~\mathrm{GeV}$ kinematics, $A_{LU}^{\rm RC}$ lies substantially below $A_{LU}^{\rm Born}$, consistent with $\delta > 1$ throughout. Both curves for $\cos\theta^{\ast} = +0.750$ decrease toward zero at large $Q^{2}$ but at different rates; see Figure ~\ref{fig:s22_ra_q2} and the associated discussion.}
  \label{fig:s22_alu_q2}
\end{figure*}

\begin{figure*}[tbp]
  \centering
  \includegraphics[width=0.8\textwidth]{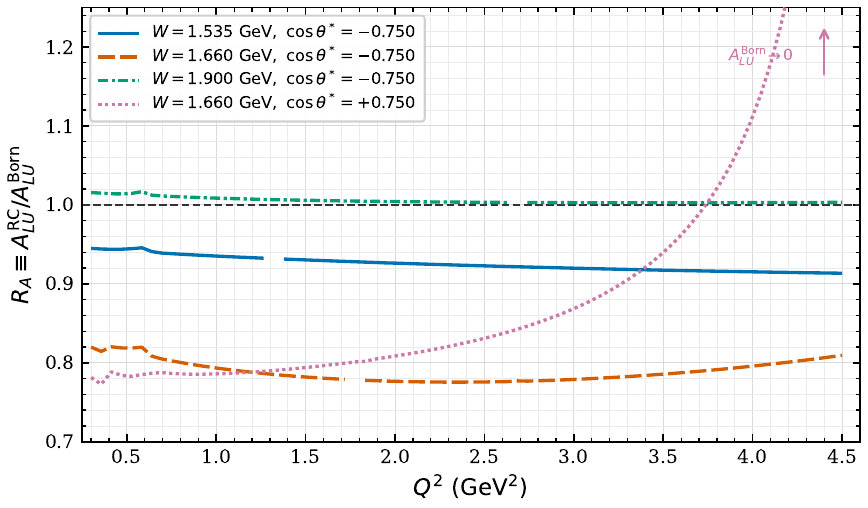}
  \caption{Asymmetry ratio $R_A \equiv A_{LU}^{\rm RC}/A_{LU}^{\rm Born}$ vs.\ $Q^{2}$ at $\phi^{\ast} = 90^{\circ}$, $v_{\mathrm{cut}} = 0.166~\mathrm{GeV}^{2}$, for four kinematic combinations (see legend). Dashed horizontal line: $R_A = 1$ (no RC effect on the asymmetry). At $W = 1.535~\mathrm{GeV}$, $R_A$ decreases slowly from $0.945$ to $0.913$. At $W = 1.900~\mathrm{GeV}$, $R_A$ lies within ${\sim}\,2\%$ of unity and crosses $R_A = 1$ near $Q^{2} \approx 3.4~\mathrm{GeV}^{2}$. At $W = 1.660~\mathrm{GeV}$ (backward), $R_A \approx 0.775$--$0.820$ and is nearly $Q^{2}$-independent. At $W = 1.660~\mathrm{GeV}$ (forward), $R_A \approx 0.78$--$0.82$ for $Q^{2} \lesssim 2~\mathrm{GeV}^{2}$ and then diverges; the curve is clipped at $R_A = 1.25$ and the divergence is indicated by the arrow. Both $A_{LU}^{\rm RC}$ and $A_{LU}^{\rm Born}$ approach zero at these forward-angle, large-$Q^{2}$ kinematics, but the radiative integral in $A_{LU}^{\rm RC}$ samples neighboring phase-space points where $A_{LU}^{\rm Born}$ has not yet reached its zero, so the numerator vanishes more slowly than the denominator and $R_A$ diverges. This is a kinematic artifact of the ratio definition near a zero of the denominator; $R_A$ is not a meaningful diagnostic in this region.}
  \label{fig:s22_ra_q2}
\end{figure*}

% ------------------------------------------------------------------
Figures~\ref{fig:s22_delta_q2}--\ref{fig:s22_ra_q2} show $\delta$, $A_{LU}$, and $R_A$ as functions of $Q^{2}$ at fixed $\phi^{\ast} = 90^{\circ}$, $v_{\mathrm{cut}} = 0.166~\mathrm{GeV}^{2}$, for four kinematic combinations: the N(1535) resonance ($W = 1.535~\mathrm{GeV}$, $\cos\theta^{\ast} = -0.750$), the $\delta$-peak at a backward angle ($W = 1.660~\mathrm{GeV}$, $\cos\theta^{\ast} = -0.750$), the post-peak plateau ($W = 1.900~\mathrm{GeV}$, $\cos\theta^{\ast} = -0.750$), and the $\delta$-peak at a forward angle ($W = 1.660~\mathrm{GeV}$, $\cos\theta^{\ast} = +0.750$). Each curve contains 75 evenly spaced points from $Q^{2} = 0.30$ to $4.50~\mathrm{GeV}^{2}$.

Three isolated points produce undefined output at $(W,Q^{2}) = (1.535, 1.322)$, $(1.660, 1.776)$, and $(1.900, 2.684)~\mathrm{GeV}^{2}$, appearing as small single-point gaps in the corresponding curves. These locations shifted relative to a prior 50-point run, ruling out a grid-density effect; they are instead isolated nodes of the EtaMAID-2023 multipole interpolation grid where the spline evaluation returns a non-finite value. Each gap spans a single grid step (${\sim}\,0.057~\mathrm{GeV}^{2}$) and does not affect the surrounding curves.

All four curves exhibit a localized non-monotonic feature near $Q^{2} \approx 0.55$--$0.70~\mathrm{GeV}^{2}$, visible as an inflection or local extreme in $\delta$, a sharp dip in $A_{LU}$, and a small kink in $R_A$. This feature persists at 75-point resolution and is therefore physical. Notably, it appears coherently across all four kinematic combinations despite their different $W$ and $\cos\theta^{\ast}$ values, pointing to a $Q^{2}$-driven threshold or multipole interference in the EtaMAID-2023 tables rather than a hadronic amplitude artifact specific to any one resonance.

The $\delta$ scan (Figure ~\ref{fig:s22_delta_q2}) separates cleanly into three regimes. At the N(1535) resonance ($W = 1.535~\mathrm{GeV}$), $\delta < 1$ is nearly flat throughout, ranging from $\delta \approx 0.827$ at $Q^{2} \approx 0.58~\mathrm{GeV}^{2}$ to $\delta \approx 0.848$ at $Q^{2} = 4.50~\mathrm{GeV}^{2}$; RC reduces the cross section by roughly $15$--$17\%$ uniformly across the accessible $Q^{2}$ range. At the post-peak plateau ($W = 1.900~\mathrm{GeV}$), $\delta$ is similarly below unity and rises slowly from $0.864$ to $0.911$ over the same range. At the $\delta$-peak ($W = 1.660~\mathrm{GeV}$, backward), $\delta > 1$ throughout and increases monotonically from $1.050$ to $1.266$, so that the RC inflation of the cross section grows from $5\%$ at $Q^{2} = 0.30~\mathrm{GeV}^{2}$ to $27\%$ at $Q^{2} = 4.50~\mathrm{GeV}^{2}$. The forward-angle companion ($W = 1.660~\mathrm{GeV}$, $\cos\theta^{\ast} = +0.750$) shows the opposite $Q^{2}$ trend: $\delta$ falls from $1.561$ at $Q^{2} = 0.30~\mathrm{GeV}^{2}$ toward $1.209$ at high $Q^{2}$, with the non-monotonic bump near $Q^{2} \approx 0.64~\mathrm{GeV}^{2}$ interrupting the smooth decrease. The two $W = 1.660~\mathrm{GeV}$ curves converge and cross near $Q^{2} \approx 2.5~\mathrm{GeV}^{2}$, above which the backward-angle curve overtakes the forward-angle one.

The asymmetry scan (Figure ~\ref{fig:s22_alu_q2}) shows that $A_{LU}^{\rm Born}$ and $A_{LU}^{\rm RC}$ track each other closely at $W = 1.535~\mathrm{GeV}$, with both curves being small (${\lesssim}\,30\%$) and nearly coincident across the full $Q^{2}$ range. At $W = 1.900~\mathrm{GeV}$, the asymmetry is large and $Q^{2}$-dependent, starting near $185\%$ at low $Q^{2}$, falling sharply through the bump region, and decreasing smoothly to ${\sim}\,85\%$ at $Q^{2} = 4.50~\mathrm{GeV}^{2}$; the RC and Born curves are nearly indistinguishable throughout. At $W = 1.660~\mathrm{GeV}$ (both angles), the RC–Born gap is substantial: $A_{LU}^{\rm RC}$ lies well below $A_{LU}^{\rm Born}$, consistent with $\delta > 1$ throughout. Notably, both $A_{LU}^{\rm RC}$ and $A_{LU}^{\rm Born}$ decrease toward zero at large $Q^{2}$ for $\cos\theta^{\ast} = +0.750$; however, as discussed below, the two quantities approach zero at different rates.

The asymmetry ratio (Figure ~\ref{fig:s22_ra_q2}) provides a cleaner separation of the RC effect from the hadronic amplitude. At $W = 1.535~\mathrm{GeV}$, $R_A$ decreases slowly from $0.945$ to $0.913$, indicating a mild but increasing RC suppression with $Q^{2}$. At $W = 1.900~\mathrm{GeV}$, $R_A$ lies just above unity at low $Q^{2}$ ($R_A \approx 1.017$ near $Q^{2} \approx 0.58~\mathrm{GeV}^{2}$), crosses $R_A = 1$ near $Q^{2} \approx 3.4~\mathrm{GeV}^{2}$, and remains within ${\sim}\,2\%$ of unity throughout; the sign of the RC correction to the asymmetry changes within the accessible experimental range. At $W = 1.660~\mathrm{GeV}$ (backward), $R_A$ is approximately flat at $0.775$--$0.820$, with a weak tendency to increase at high $Q^{2}$; RC suppression of the asymmetry is ${\sim}\,20\%$ and only mildly $Q^{2}$-dependent.

At $W = 1.660~\mathrm{GeV}$ (forward), $R_A \approx 0.78$--$0.82$ for $Q^{2} \lesssim 2~\mathrm{GeV}^{2}$, it then rises rapidly, exiting the plotted range above $Q^{2} \approx 4.2~\mathrm{GeV}^{2}$. Although both $A_{LU}^{\rm RC}$ and $A_{LU}^{\rm Born}$ are decreasing toward zero at these kinematics (visible in Figure ~\ref{fig:s22_alu_q2}), they do so at different rates. The RC correction to an asymmetry is not multiplicative: $A_{LU}^{\rm RC}$ is computed by integrating the Born amplitude over neighboring kinematics, weighted by the real-emission radiator function, whereas $A_{LU}^{\rm Born}$ is evaluated at the single nominal phase-space point. When $A_{LU}^{\rm Born}$ possesses a kinematic near-zero, as it does at forward angles and large $Q^{2}$, the Born amplitude varies rapidly in the surrounding phase space. The radiative integral therefore samples points where $A_{LU}^{\rm Born}$ has not yet reached its zero, so $A_{LU}^{\rm RC}$ vanishes more slowly than $A_{LU}^{\rm Born}$ and $R_A \to \infty$. This is not a physically large RC effect on the asymmetry; it is a kinematic artifact of the ratio definition near a zero of the denominator, and $R_A$ is not a meaningful diagnostic in this region.

\clearpage

% ------------------------------------------------------------------
\subsubsection*{\texorpdfstring{S2.3\quad $\phi^{\ast}$ dependence:
$Q^{2}$ and $W$ scans}{S2.3 phi* dependence: Q2 and W scans}}%
\label{SM:S2.3}
% ------------------------------------------------------------------
At $W = 1.660~\mathrm{GeV}$ the $\delta(W)$ profile reaches its local
maximum (cf.\ Figure ~\ref{fig:delta_W_multiQ2}).
Each figure below pairs $\delta(\phi^{\ast})$ and $A_{LU}(\phi^{\ast})$
at the same kinematics for direct comparison.

\begin{figure*}[tbp]
  \centering
  \begin{minipage}[t]{0.48\textwidth}
    \centering
    \includegraphics[width=\linewidth]{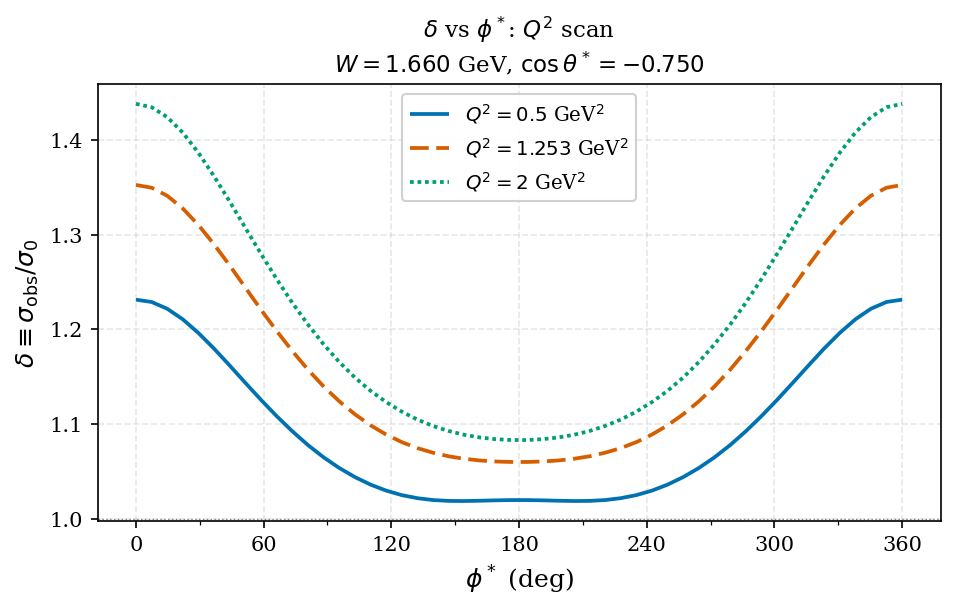}\\
    {\small (a)}
  \end{minipage}
  \hfill
  \begin{minipage}[t]{0.48\textwidth}
    \centering
    \includegraphics[width=\linewidth]{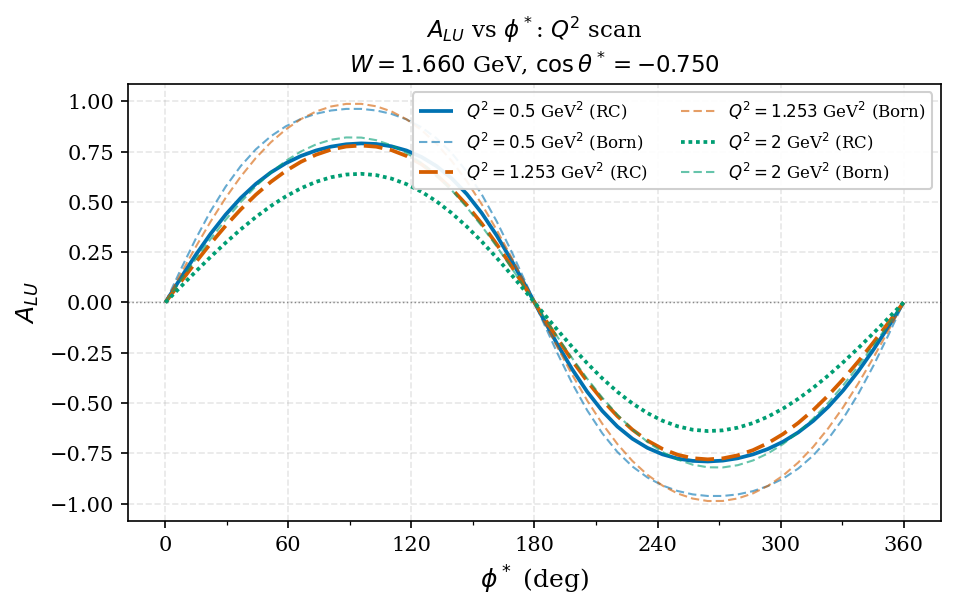}\\
    {\small (b)}
  \end{minipage}
  \caption{%
    $Q^{2}$ scan at $W = 1.660~\mathrm{GeV}$,
    $\cos\theta^{\ast} = -0.750$, $v_{\mathrm{cut}} = 0.166~\mathrm{GeV}^{2}$,
    for $Q^{2} = 0.500$, $1.253$, $2.000~\mathrm{GeV}^{2}$ (see legend).
    (a)~$\delta$ vs.\ $\phi^{\ast}$: cosine-like modulation with $\delta > 1$
    throughout; amplitude and offset increase with $Q^{2}$.
    Dashed horizontal line: $\delta = 1$.
    (b)~$A_{LU}(\phi^{\ast})$, RC (solid) and Born (dashed):
    $\sin\phi^{\ast}$ modulation; RC suppresses the amplitude by
    ${\sim}\,18$--$22\%$, with larger suppression at higher $Q^{2}$.%
  }
  \label{fig:s23_phi_q2scan}
\end{figure*}

In the $Q^{2}$ scan (Figure ~\ref{fig:s23_phi_q2scan}), all three curves
have $\delta > 1$ with a cosine-like modulation whose amplitude and
offset grow with $Q^{2}$: the ranges are $[1.02, 1.23]$, $[1.06, 1.35]$,
and $[1.08, 1.44]$ at $Q^{2} = 0.500$, $1.253$, and
$2.000~\mathrm{GeV}^{2}$, respectively.
RC suppresses the $A_{LU}$ amplitude by ${\sim}\,18$--$22\%$, with
slightly larger suppression at higher $Q^{2}$.

\begin{figure*}[tbp]
  \centering
  \begin{minipage}[t]{0.48\textwidth}
    \centering
    \includegraphics[width=\linewidth]{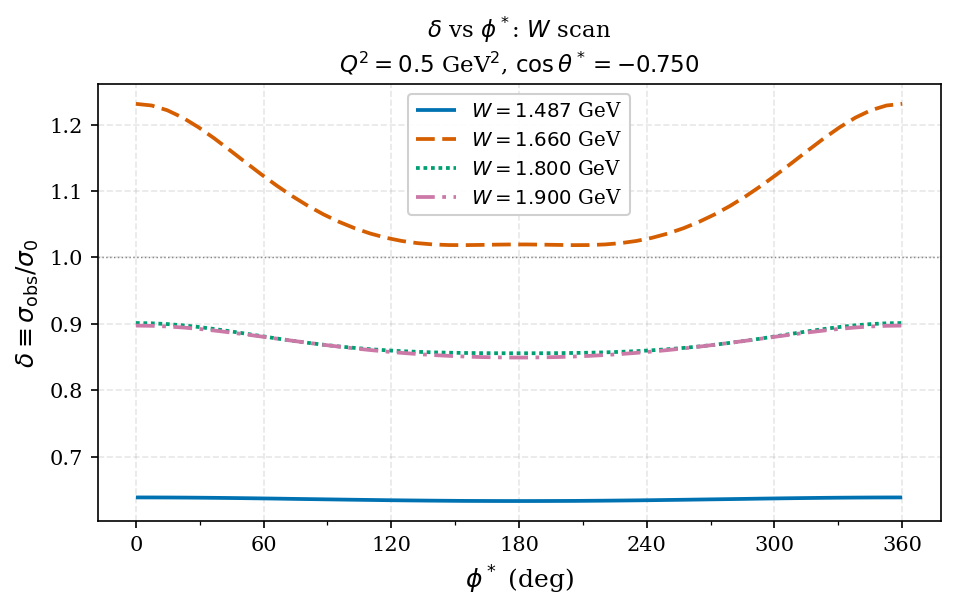}\\
    {\small (a)}
  \end{minipage}
  \hfill
  \begin{minipage}[t]{0.48\textwidth}
    \centering
    \includegraphics[width=\linewidth]{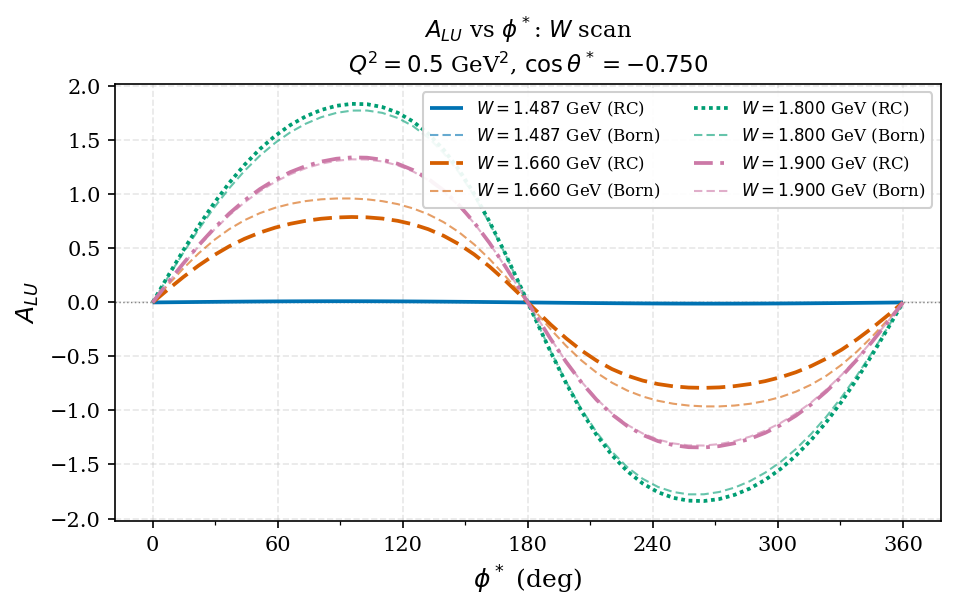}\\
    {\small (b)}
  \end{minipage}
  \caption{%
    $W$ scan at $Q^{2} = 0.500~\mathrm{GeV}^{2}$,
    $\cos\theta^{\ast} = -0.750$, $v_{\mathrm{cut}} = 0.166~\mathrm{GeV}^{2}$,
    for $W = 1.487$, $1.660$, $1.800$, $1.900~\mathrm{GeV}$ (see legend).
    (a)~$\delta$ vs.\ $\phi^{\ast}$: near-threshold ($W = 1.487~\mathrm{GeV}$)
    is nearly flat at $\delta \approx 0.63$--$0.64$; at $W = 1.660~\mathrm{GeV}$,
    $\delta > 1$ with the largest modulation; at $W = 1.800$ and
    $1.900~\mathrm{GeV}$, $\delta < 1$ with moderate modulation.
    Dashed horizontal line: $\delta = 1$.
    (b)~$A_{LU}(\phi^{\ast})$, RC (solid) and Born (dashed):
    at $W = 1.487~\mathrm{GeV}$ the two nearly overlap
    ($|A_{LU}| \lesssim 0.012$); at $W = 1.660~\mathrm{GeV}$, RC lies
    ${\sim}\,18\%$ below Born; at $W = 1.800$ and $1.900~\mathrm{GeV}$,
    RC slightly exceeds Born ($R_A > 1$).%
  }
  \label{fig:s23_phi_wscan}
\end{figure*}

In the $W$ scan (Figure ~\ref{fig:s23_phi_wscan}), the four curves span
qualitatively different RC regimes: near-threshold ($W = 1.487~\mathrm{GeV}$,
$\delta \approx 0.63$--$0.64$, nearly flat), the $\delta$ peak
($W = 1.660~\mathrm{GeV}$, $\delta > 1$), and the post-peak plateau
($W = 1.800$ and $1.900~\mathrm{GeV}$, $\delta < 1$).
The RC--Born gap in $A_{LU}$ changes sign across $W$: at
$W = 1.660~\mathrm{GeV}$, RC lies below Born; at $W = 1.800$ and
$1.900~\mathrm{GeV}$, RC slightly exceeds Born, consistent with
$R_A > 1$ in Figure ~\ref{fig:RA_W_multiQ2}.

\clearpage

% ------------------------------------------------------------------
\subsubsection*{\texorpdfstring{S2.4\quad Asymmetry ratio $R_A$
versus $\phi^{\ast}$}{S2.4 Asymmetry ratio R\_A versus phi*}}%
\label{SM:S2.4}
% ------------------------------------------------------------------
The ratio $R_A \equiv A_{LU}^{\rm RC}/A_{LU}^{\rm Born}$ isolates the
RC effect on the asymmetry independently of the hadronic model amplitude.
It is undefined near $\phi^{\ast} = 0^{\circ}$ and $180^{\circ}$ where
$A_{LU}^{\rm Born} \to 0$.

\begin{figure*}[tbp]
  \centering
  \begin{minipage}[t]{0.48\textwidth}
    \centering
    \includegraphics[width=\linewidth]{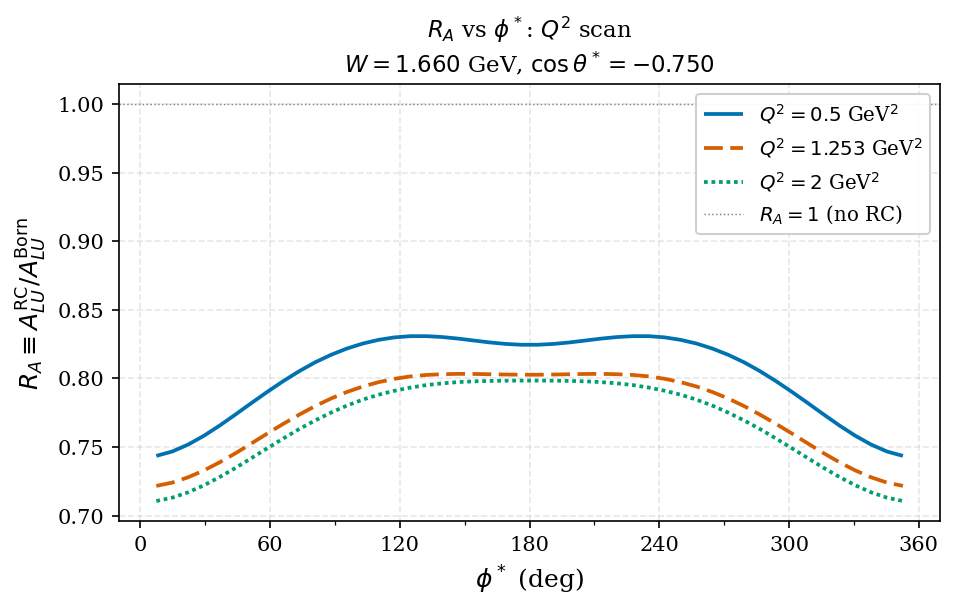}\\
    {\small (a)}
  \end{minipage}
  \hfill
  \begin{minipage}[t]{0.48\textwidth}
    \centering
    \includegraphics[width=\linewidth]{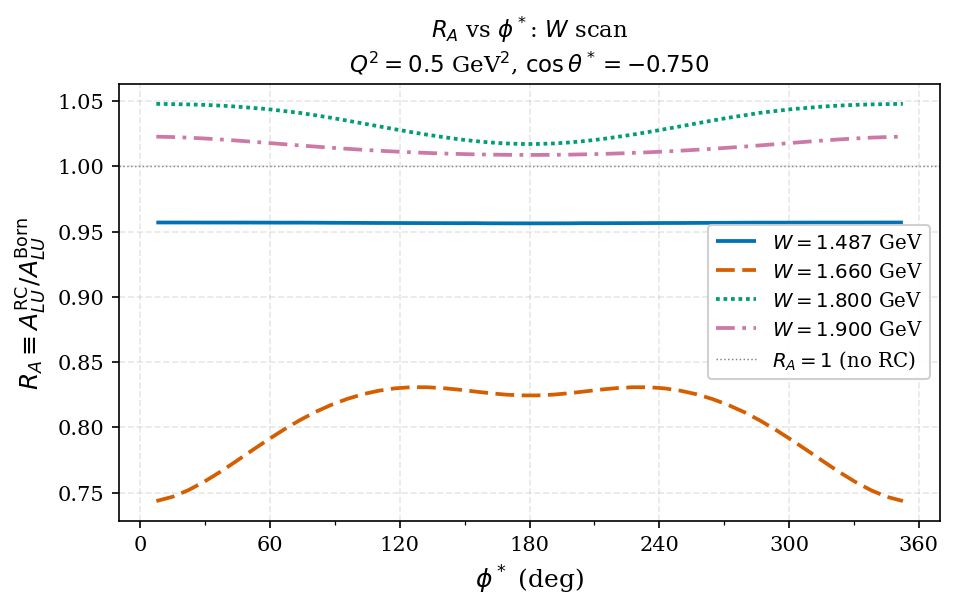}\\
    {\small (b)}
  \end{minipage}
  \caption{%
    $R_A \equiv A_{LU}^{\rm RC}/A_{LU}^{\rm Born}$ vs.\ $\phi^{\ast}$,
    $v_{\mathrm{cut}} = 0.166~\mathrm{GeV}^{2}$.
    $R_A$ is undefined near $\phi^{\ast} = 0^{\circ}$ and $180^{\circ}$
    where $A_{LU}^{\rm Born} \to 0$.
    Dashed horizontal line: $R_A = 1$ (no RC effect on asymmetry).
    (a)~$Q^{2}$ scan at $W = 1.660~\mathrm{GeV}$,
    $\cos\theta^{\ast} = -0.750$, for $Q^{2} = 0.500$, $1.253$,
    $2.000~\mathrm{GeV}^{2}$: $R_A \approx 0.71$--$0.83$ throughout;
    suppression deepens with $Q^{2}$.
    (b)~$W$ scan at $Q^{2} = 0.500~\mathrm{GeV}^{2}$,
    $\cos\theta^{\ast} = -0.750$, for $W = 1.487$, $1.660$, $1.800$,
    $1.900~\mathrm{GeV}$: $R_A$ transitions from slight suppression at
    $W = 1.487~\mathrm{GeV}$ to deepest suppression at
    $W = 1.660~\mathrm{GeV}$, then to slight enhancement at $W = 1.800$
    and $1.900~\mathrm{GeV}$.%
  }
  \label{fig:s24_ra_phi}
\end{figure*}

Figure~\ref{fig:s24_ra_phi} presents the $Q^{2}$ and $W$ scans side by
side.
In the $Q^{2}$ scan (left panel), all curves lie in the range
$R_A \approx 0.71$--$0.83$, with a weak $\phi^{\ast}$ modulation
reaching its minimum near $\phi^{\ast} = 90^{\circ}$ and $270^{\circ}$;
suppression deepens slightly with $Q^{2}$.
In the $W$ scan (right panel), the four curves span qualitatively
different regimes: $R_A \approx 0.957$ at $W = 1.487~\mathrm{GeV}$
(nearly flat, slight suppression); $R_A \in [0.74, 0.83]$ at
$W = 1.660~\mathrm{GeV}$ (deepest suppression); $R_A > 1$ at
$W = 1.800$ and $1.900~\mathrm{GeV}$ (${\sim}\,1.009$--$1.048$), a
sign reversal consistent with the $\delta < 1$ regime and the $R_A$
plateau in Figure ~\ref{fig:RA_W_multiQ2}.

\clearpage

% ================================================================
% SM S3 — sin phi* moment of the beam-spin asymmetry
% ================================================================
\subsection*{\texorpdfstring{S3. $\sin\phi^{\ast}$ moment of the beam-spin asymmetry}{S3. sin(phi*) moment of the beam-spin asymmetry}}\label{SM:S3}
The beam-spin asymmetry $A_{LU}$ for $\eta$ electroproduction is dominated by its $\sin\phi^{\ast}$ harmonic, which carries the leading-twist interference between longitudinal and transverse virtual-photon amplitudes. To quantify the effect of radiative corrections at the level of this observable, we extract the $\sin\phi^{\ast}$ moment from the EXCLURAD model output and form the ratio of the corrected to uncorrected amplitudes across the full kinematic grid. All results in this section use the EtaMAID multipoles described in the main text; no experimental data are included.

\paragraph{Definition of the model asymmetries.}
The Born and radiatively corrected asymmetries reported by EXCLURAD at each kinematic point $(W,\,Q^{2},\,\cos\theta^{\ast},\,\phi^{\ast})$ are stored as percentages. We convert to fractional asymmetries via
\[
A_{LU}^{\mathrm{Born}}(\phi^{\ast}) = \frac{\texttt{asb}(\phi^{\ast})}{100}\,, \qquad A_{LU}^{\mathrm{RC}}(\phi^{\ast}) = \frac{\texttt{as}(\phi^{\ast})}{100}\,.
\]
No beam-polarization factor or sign convention adjustment is applied.

\paragraph{Extraction of the $\sin\phi^{\ast}$ moment.}
For each bin in $(W,\,Q^{2},\,\cos\theta^{\ast})$ we fit the one-parameter form
\begin{equation}
    A_{LU}(\phi^{\ast}) \;=\; A_{LU}^{\sin\phi^{\ast}}\,\sin\!\bigl(\phi^{\ast}\bigr)\,,
    \label{eq:sine_fit_s3}
\end{equation}
to the $N_{\phi^{\ast}}$ azimuthal points (typically 30) using a \textsc{Root}
\texttt{TGraph} fit with $\phi^{\ast}$ in degrees (converted internally to radians).
The fit returns $A_{LU}^{\sin\phi^{\ast}}$ and its uncertainty from the covariance
matrix. Because the EXCLURAD output is model-generated rather than experimental data,
no per-point statistical errors are assigned; \textsc{Root} computes the parameter
uncertainty from the curvature of the residual sum of squares with respect to the
amplitude parameter. This uncertainty measures how well a single-harmonic sine function
describes the model output at that kinematic bin, not a physical statistical
uncertainty. The ratio $|A_{LU}^{\sin\phi^{\ast}}|/\sigma_{A}$ used in the quality
cut should therefore be read as a fit-quality diagnostic: bins where this ratio falls
below~3 are excluded because the sine fit is poorly constrained relative to the
fitting residual, not because of low statistical significance in the frequentist
sense. Out of 28\,417 kinematic bins, 26\,887 pass these quality cuts.

\begin{figure}[!htbp]
    \centering
    \includegraphics[width=0.85\columnwidth]{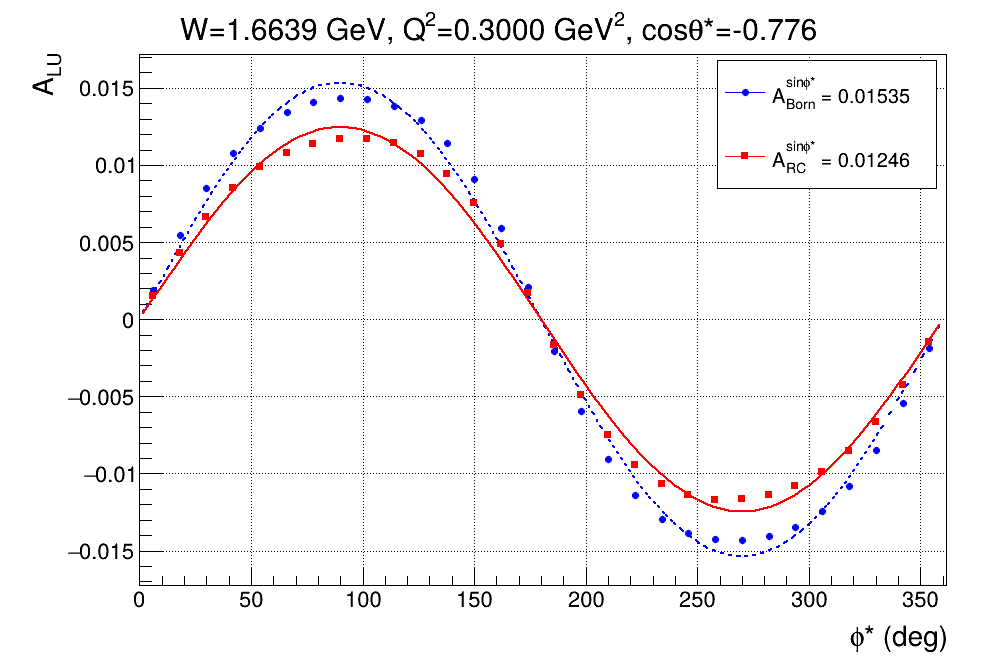}
    \caption{Beam-spin asymmetry $A_{LU}$ vs.\ $\phi^{\ast}$ at $W = 1.6639$~GeV, $Q^{2} = 0.300$~GeV$^{2}$, $\cos\theta^{\ast} = -0.776$. Blue circles: Born; red squares: radiatively corrected. Curves show the $\sin\phi^{\ast}$ fits (Eq.~\ref{eq:sine_fit_s3}), with the extracted amplitudes indicated in the legend. The dotted line marks $A_{LU} = 0$.}
    \label{fig:s3_test_bin}
\end{figure}

Figure~\ref{fig:s3_test_bin} shows a representative fit at $W = 1.6639$~GeV, $Q^{2} = 0.300$~GeV$^{2}$, $\cos\theta^{\ast} = -0.776$, where the Born and RC asymmetries both follow the expected $\sin\phi^{\ast}$ dependence. The fits yield $A_{LU,\,\mathrm{Born}}^{\sin\phi^{\ast}} = 0.01535$ and $A_{LU,\,\mathrm{RC}}^{\sin\phi^{\ast}} = 0.01246$, giving a ratio of $0.812$: the RC reduces the $\sin\phi^{\ast}$ amplitude by approximately 19\% at this kinematic point, which lies near the peak of $\delta$ reported in the main text (Figs.~5 and~8).

\begin{figure}[!htbp]
    \centering
    \includegraphics[width=0.85\columnwidth]{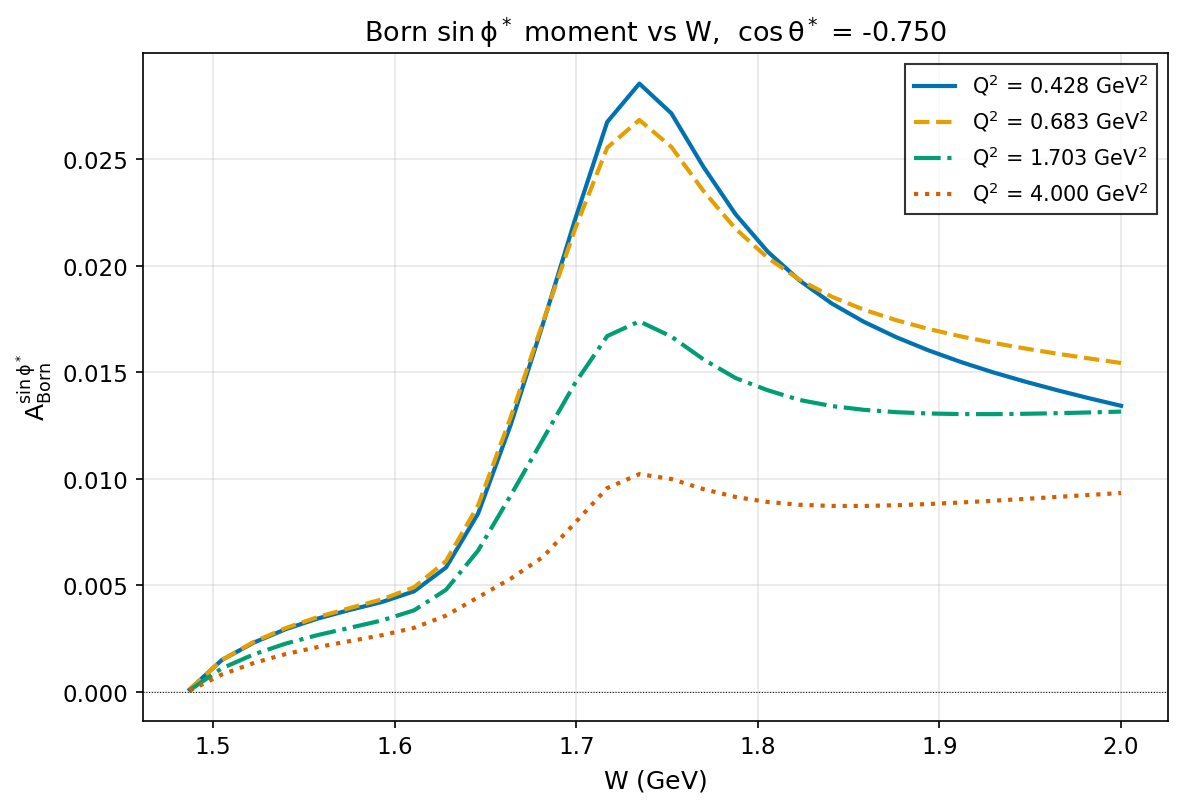}
    \caption{Born $\sin\phi^{\ast}$ moment $A_{LU,\,\mathrm{Born}}^{\sin\phi^{\ast}}$ vs.\ $W$ at $\cos\theta^{\ast} = -0.750$ for four $Q^{2}$ values: $0.428$ (blue solid), $0.683$ (orange dashed), $1.703$ (teal dot-dashed), and $4.000$~GeV$^{2}$ (vermilion dotted).}
    \label{fig:s3_born_vs_W}
\end{figure}

\begin{figure}[!htbp]
    \centering
    \includegraphics[width=0.85\columnwidth]{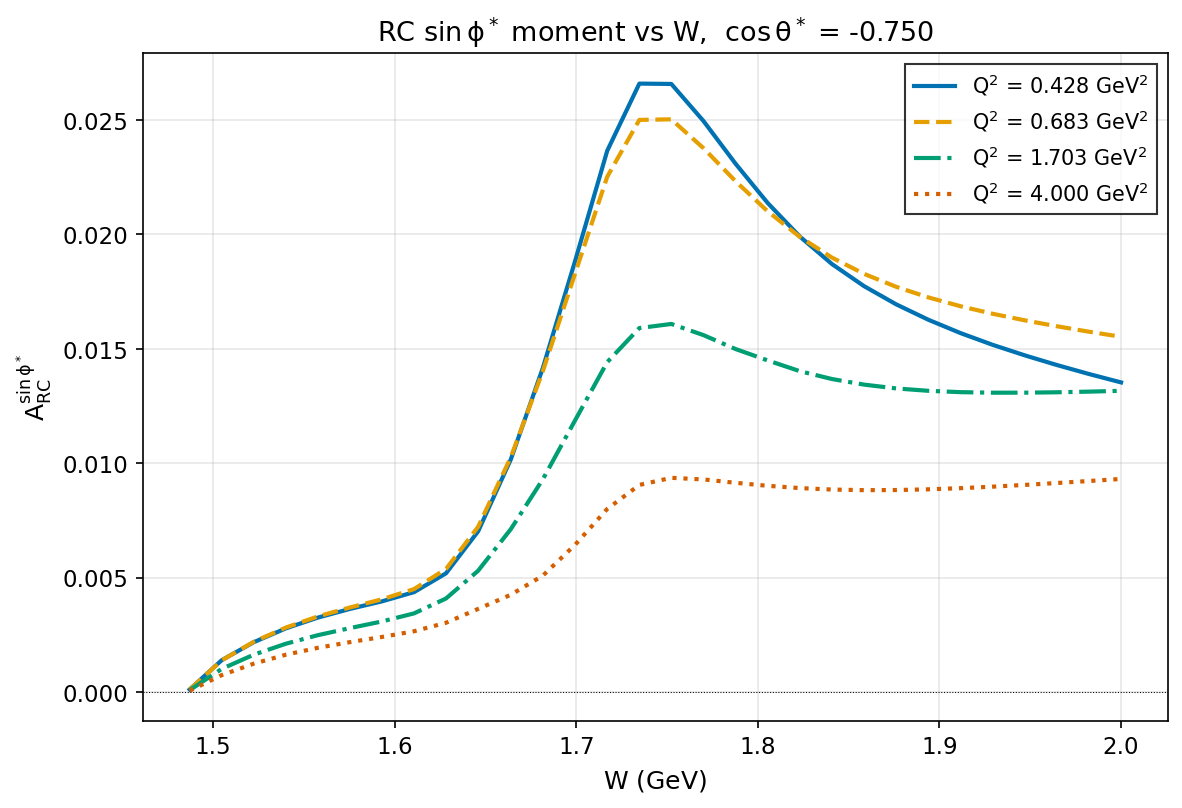}
    \caption{Same as Figure ~\ref{fig:s3_born_vs_W} but for the radiatively corrected asymmetry. The shape and $Q^{2}$ ordering are preserved; all curves are shifted downward relative to Born.}
    \label{fig:s3_rc_vs_W}
\end{figure}

\paragraph{Born and RC sine moments vs.\ $W$.}
Figures~\ref{fig:s3_born_vs_W} and~\ref{fig:s3_rc_vs_W} show $A_{LU}^{\sin\phi^{\ast}}$ as a function of $W$ at $\cos\theta^{\ast} = -0.750$ for four representative $Q^{2}$ values spanning the grid. All curves rise from near zero at the $\eta$ threshold, reach a broad maximum around $W \approx 1.70$--$1.75$~GeV in the $S_{11}(1535)$/$S_{11}(1650)$ region, and then decrease or level off toward $W = 2.0$~GeV. The amplitude falls with increasing $Q^{2}$, consistent with the $Q^{2}$ dependence of the electromagnetic transition form factors. Comparing the two figures, the RC curves preserve the resonance structure and $Q^{2}$ ordering of the Born curves but are systematically lower in magnitude.

\paragraph{Sine-moment ratio.}
To isolate the fractional RC effect, we form
\begin{equation}
    R_{A}^{\sin\phi^{\ast}}(W,\,Q^{2},\,\cos\theta^{\ast}) \;=\; \frac{A_{LU,\,\mathrm{RC}}^{\sin\phi^{\ast}}}{A_{LU,\,\mathrm{Born}}^{\sin\phi^{\ast}}}\,,
    \label{eq:sine_ratio_s3}
\end{equation}
where a value of unity indicates no RC effect, and values below unity indicate suppression by radiative photon emission. The uncertainty on $R_{A}^{\sin\phi^{\ast}}$ is propagated from the individual fit uncertainties in quadrature.

\begin{figure}[!htbp]
    \centering
    \includegraphics[width=0.85\columnwidth]{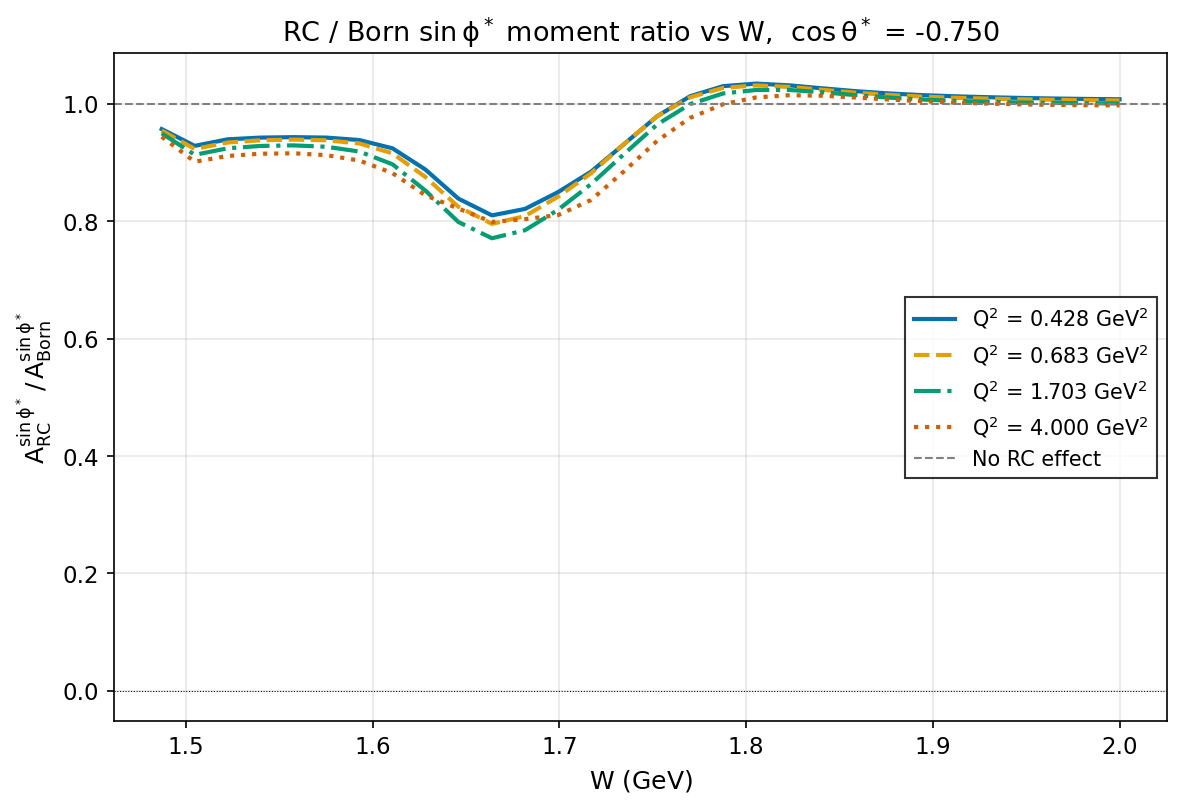}
    \caption{Ratio $R_{A}^{\sin\phi^{\ast}}$ (Eq.~\ref{eq:sine_ratio_s3}) vs.\ $W$ at $\cos\theta^{\ast} = -0.750$, for $Q^{2} = 0.428$ (blue solid), $0.683$ (orange dashed), $1.703$ (teal dot-dashed), and $4.000$~GeV$^{2}$ (vermillion dotted). The dip near $W \approx 1.66$~GeV corresponds to the peak of $\delta$ in the main text. The gray dashed line marks $R_{A}^{\sin\phi^{\ast}} = 1$ (no RC effect).}
    \label{fig:s3_ratio_back}
\end{figure}

\begin{figure}[!htbp]
    \centering
    \includegraphics[width=0.85\columnwidth]{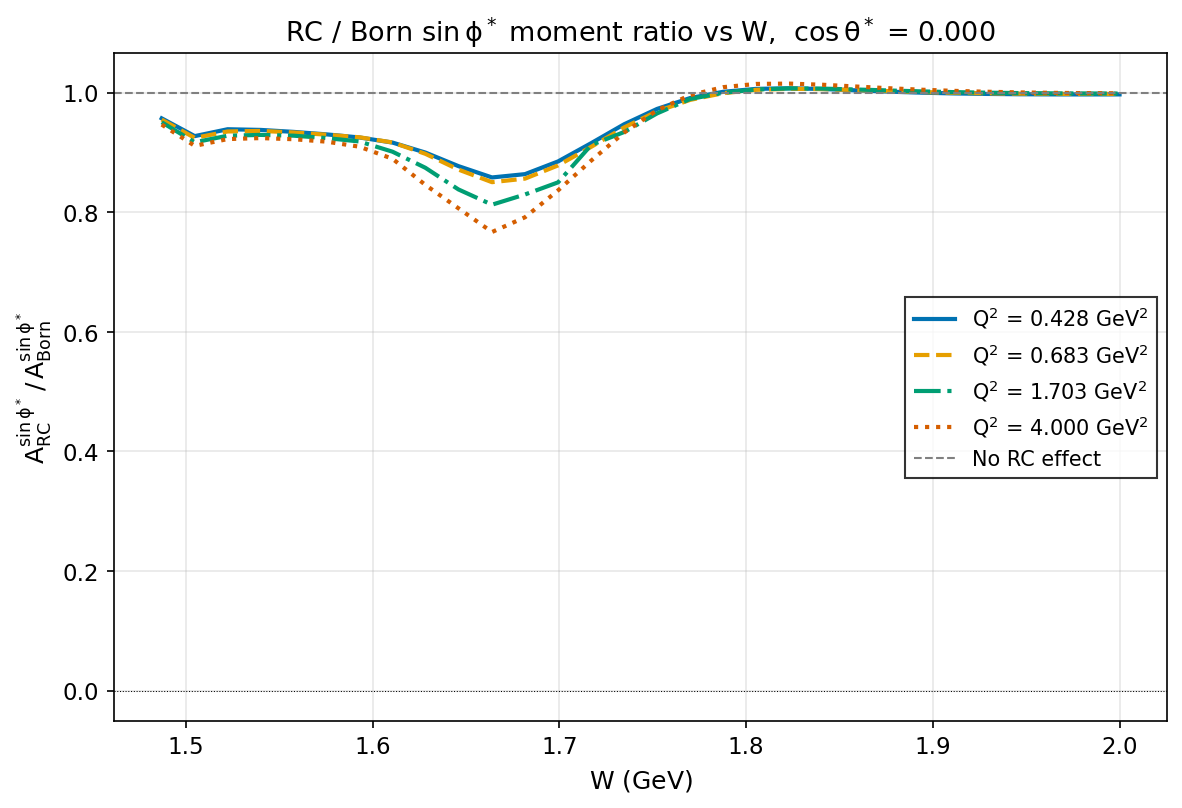}
    \caption{Same as Figure ~\ref{fig:s3_ratio_back} but at $\cos\theta^{\ast} = 0.000$. The dip depth and position are similar to the backward-angle case, confirming the weak angular dependence of $R_{A}^{\sin\phi^{\ast}}$ at central and backward angles.}
    \label{fig:s3_ratio_cent}
\end{figure}

\begin{figure}[!htbp]
    \centering
    \includegraphics[width=0.85\columnwidth]{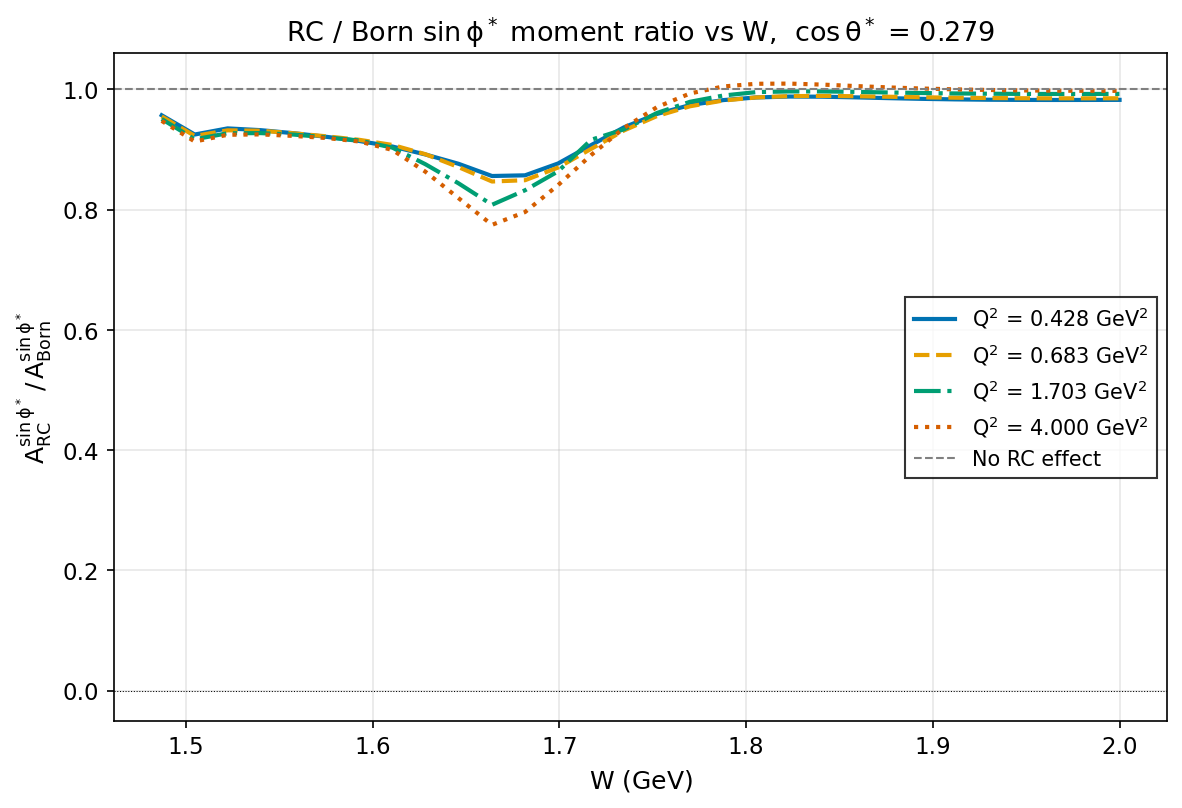}
    \caption{Same as Figure ~\ref{fig:s3_ratio_back} but at $\cos\theta^{\ast} = +0.279$. The dip remains near $W \approx 1.66$~GeV, but the $Q^{2}$ curves begin to separate. At more forward angles the ratio becomes unstable where $A_{LU,\,\mathrm{Born}}^{\sin\phi^{\ast}}$ passes near zero (see text).}
    \label{fig:s3_ratio_fwd}
\end{figure}

\paragraph{Ratio $R_{A}^{\sin\phi^{\ast}}$ vs.\ $W$ at fixed $\cos\theta^{\ast}$.}
Figures~\ref{fig:s3_ratio_back}--\ref{fig:s3_ratio_fwd} display $R_{A}^{\sin\phi^{\ast}}$ vs.\ $W$ at three $\cos\theta^{\ast}$ values. At backward and central angles ($\cos\theta^{\ast} = -0.750$ and $0.000$), the ratio is well-behaved across the full $W$ range. It starts near 0.95 at threshold, dips to approximately 0.78--0.82 around $W \approx 1.66$~GeV, then recovers to ${\sim}\,1.0$ by $W \approx 1.80$~GeV and remains there at higher $W$. The dip position coincides with the peak of $\delta$ shown in Figure ~5 of the main text, confirming that the anti-correlation between $R_{A}$ and $\delta$ observed at the full-asymmetry level persists at the sine-moment level. The $Q^{2}$ spread across the four curves is modest (a few percent), indicating that the fractional RC effect on the $\sin\phi^{\ast}$ moment depends only weakly on $Q^{2}$.

At the mildly forward angle $\cos\theta^{\ast} = +0.279$ (Figure ~\ref{fig:s3_ratio_fwd}), the overall pattern remains similar, but the $Q^{2}$ curves begin to separate more at intermediate $W$. At still more forward angles ($\cos\theta^{\ast} \gtrsim +0.5$, not shown), the ratio becomes unreliable: individual $Q^{2}$ curves develop spikes above unity near $W \approx 1.6~\mathrm{GeV}$. This instability arises because $A_{LU,\,\mathrm{Born}}^{\sin\phi^{\ast}}$ passes through or near zero in the forward-angle region, amplifying small RC shifts into large fractional changes. This is the same near-zero artifact discussed for the full asymmetry ratio $R_{A}$ in Sec.~S2.2.

\clearpage

% ================================================================
% SM S4 — Code, data, and interactive explorer
% ================================================================
\subsection*{S4. Code, data, and interactive explorer}\label{SM:S4}

The $\eta$-channel adaptation of EXCLURAD required modifications only to the non-physics layers of the code: hadron-mass constants were updated from $m_\pi$ to $m_\eta$, the I/O layer was revised to read EtaMAID-2023 multipole tables and write higher-precision output, and channel-selection flags were extended for the $\eta N$ final state. The QED radiative-correction formalism, all analytic integrands, and the bremsstrahlung integration routines are byte-for-byte identical to the CERNLIB-free distribution of Ref.~\cite{Afanasev:2002ee}. The EtaMAID-2023 multipole tables cover $W = 1.486$--$2.000$~GeV in 5~MeV steps, $Q^{2} = 0.00$--$5.01~\mathrm{GeV}^{2}$ in $0.01~\mathrm{GeV}^{2}$ steps, and angular momenta $l = 0, \ldots, 5$. Full implementation details, build notes, and input file documentation are provided in the repository.

% ------------------------------------------------------------------
\subsubsection*{\texorpdfstring{S4.1\quad Leading-log vs.\ exact comparison}{S4.1 Leading-log vs. exact comparison}}
\begin{figure*}[!htbp]
    \centering
    \includegraphics[width=0.7\textwidth]{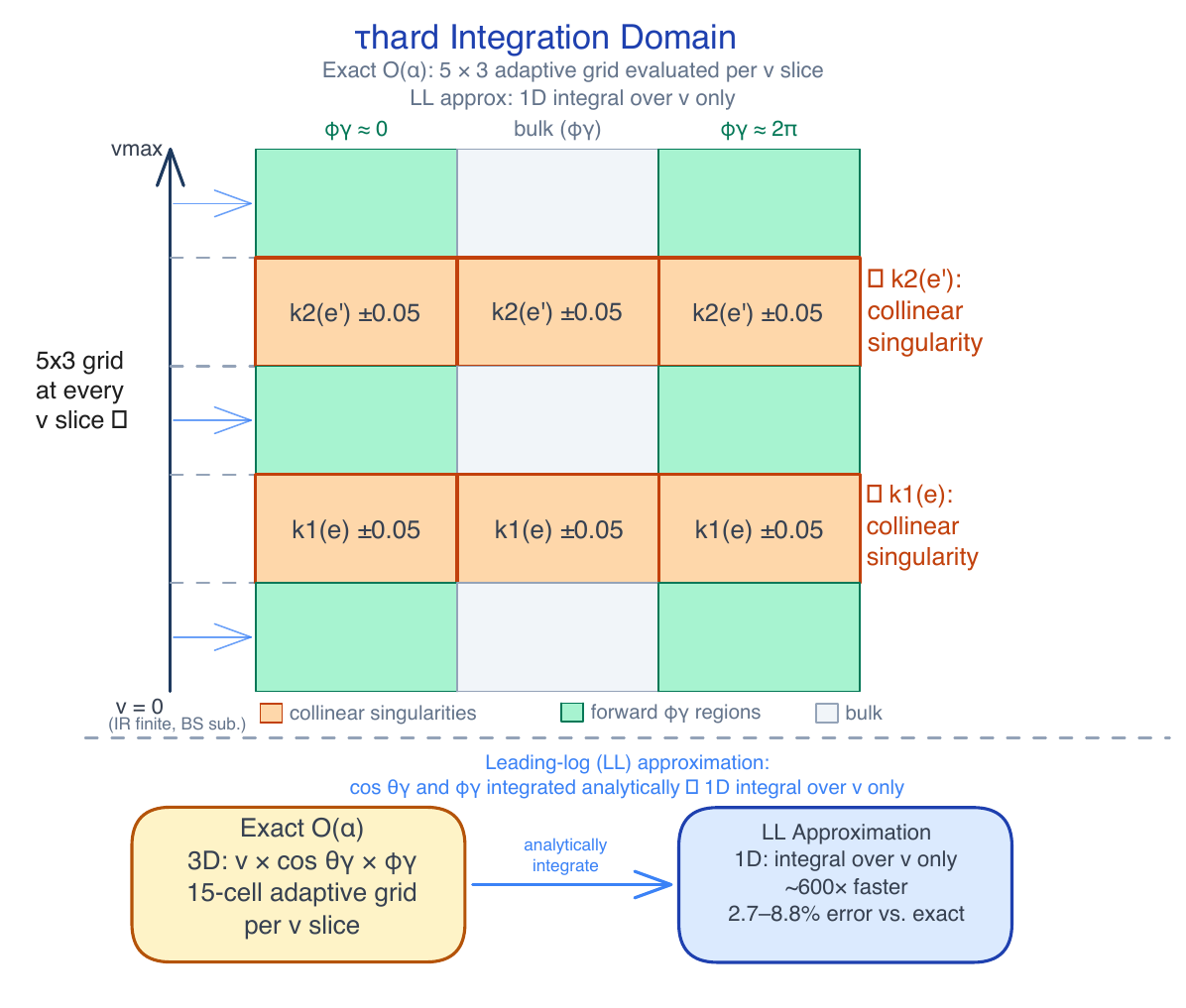}
    \caption{(a)~Integration domain for $\tau_{\rm hard}$ (Eq.~(\ref{eq:tauhard})). The $v$-axis runs from $v = 0$ (IR finite by the Bardin--Shumeiko subtraction) to $v_{\rm max}$; the sweep arrows indicate that the full $5 \times 3$ adaptive grid is re-evaluated at every $v$ slice. The grid partitions $(\cos\theta_\gamma, \phi_\gamma)$ space: orange rows bracket the collinear singularities at $\cos\theta_\gamma = \cos\theta_1$ and $\cos\theta_\gamma = \cos\theta_2$ with $\pm 0.05$ windows; green columns bracket the forward region near $\phi_\gamma = 0$ and $2\pi$; gray cells are the bulk. (b)~In the leading-log (LL) approximation, the $\cos\theta_\gamma$ and $\phi_\gamma$ integrals are performed analytically, collapsing the full three-dimensional domain to a single one-dimensional integral over $v$; the LL mode is approximately $600\times$ faster per kinematic point and deviates from the exact result by $2.7 - 8.8\%$ over the kinematic range of Sec.~\ref{sec:results}.}
    \label{fig:phasespace}
\end{figure*}

\begin{figure*}[!htbp]
    \centering
    \includegraphics[width=\textwidth]{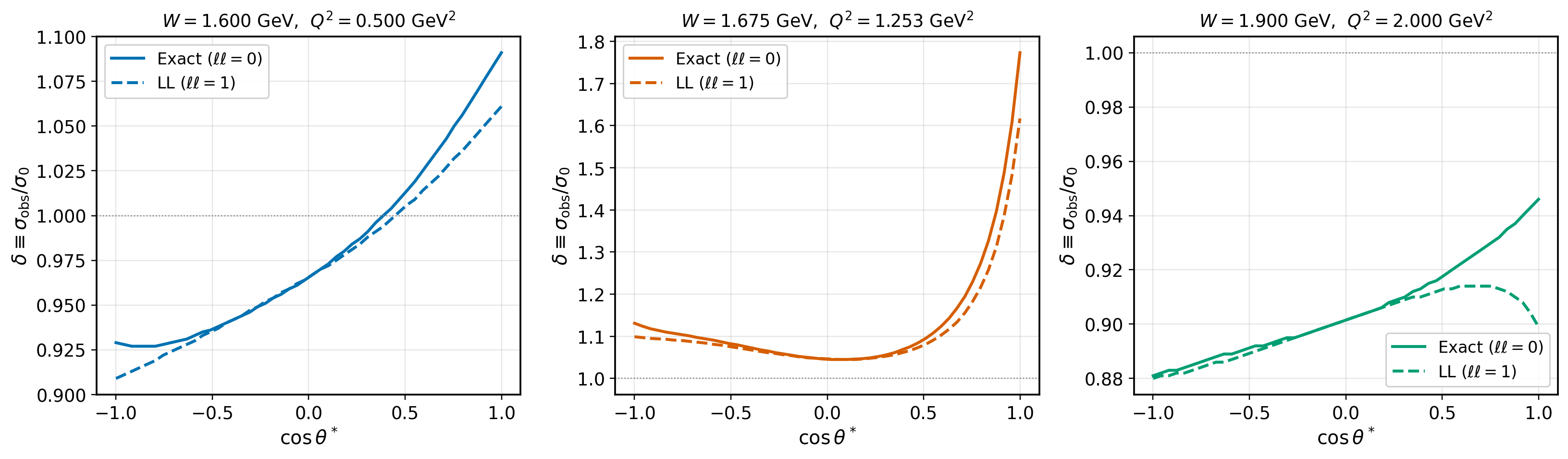}
    \caption{Leading-log (\texttt{ll\,=\,1}, dashed) vs.\ exact (\texttt{ll\,=\,0}, solid) correction factor $\delta \equiv \sigma_{\rm obs}/\sigma_0$ as a function of $\cos\theta^{\ast}$ at $\phi^{\ast} = 90^{\circ}$, for three $(W, Q^{2})$ bins (left to right): $(1.600~\mathrm{GeV},\; 0.500~\mathrm{GeV}^{2})$, $(1.675~\mathrm{GeV},\; 1.253~\mathrm{GeV}^{2})$, and $(1.900~\mathrm{GeV},\; 2.000~\mathrm{GeV}^{2})$. Both curves are extracted from a single exact-mode run; see text. The dotted horizontal line marks $\delta = 1$. The $y$-axes differ between panels owing to the large variation of $\delta$ with kinematics. $v_{\mathrm{cut}} = 0.166~\mathrm{GeV}^{2}$.}
    \label{fig:s1_ll_vs_exact}
\end{figure*}

% ------------------------------------------------------------------
EXCLURAD provides two bremsstrahlung modes: exact $O(\alpha)$ (flag \texttt{ll\,=\,0}), which evaluates a three-dimensional integral over the real-photon phase space, and a leading-log approximation (\texttt{ll\,=\,1}), which reduces to a one-dimensional integral over the inelasticity $v$. All published RC results in this work use the exact mode.
Figure~\ref{fig:phasespace} shows the integration domain for the exact mode and how the LL approximation collapses it to a single integral over $v$.
Figure~\ref{fig:s1_ll_vs_exact} compares the two modes as a function of $\cos\theta^{\ast}$ at $\phi^{\ast} = 90^{\circ}$ for three representative $(W, Q^{2})$ bins; both curves are extracted from a single exact-mode run, so the model input is identical by construction.

The leading-log approximation tracks the exact mode to within $2.7\%$ at $(W, Q^{2}) = (1.600~\mathrm{GeV}, 0.500~\mathrm{GeV}^{2})$ and $5.0\%$ at $(1.900~\mathrm{GeV}, 2.000~\mathrm{GeV}^{2})$. The largest discrepancy, $8.8\%$, occurs at $(1.675~\mathrm{GeV}, 1.253~\mathrm{GeV}^{2})$ near $\cos\theta^{\ast} \to +1$, where the hadronic model input varies rapidly. The leading-log mode is approximately $605\times$ times faster per kinematic point and is suitable for survey calculations where percent-level accuracy suffices.

% ------------------------------------------------------------------
\subsubsection*{\texorpdfstring{S4.2\quad Browser-based visualization}{S4.2 Browser-based visualization}}
\begin{figure*}[!htbp]
    \centering
    \includegraphics[width=0.85\textwidth]{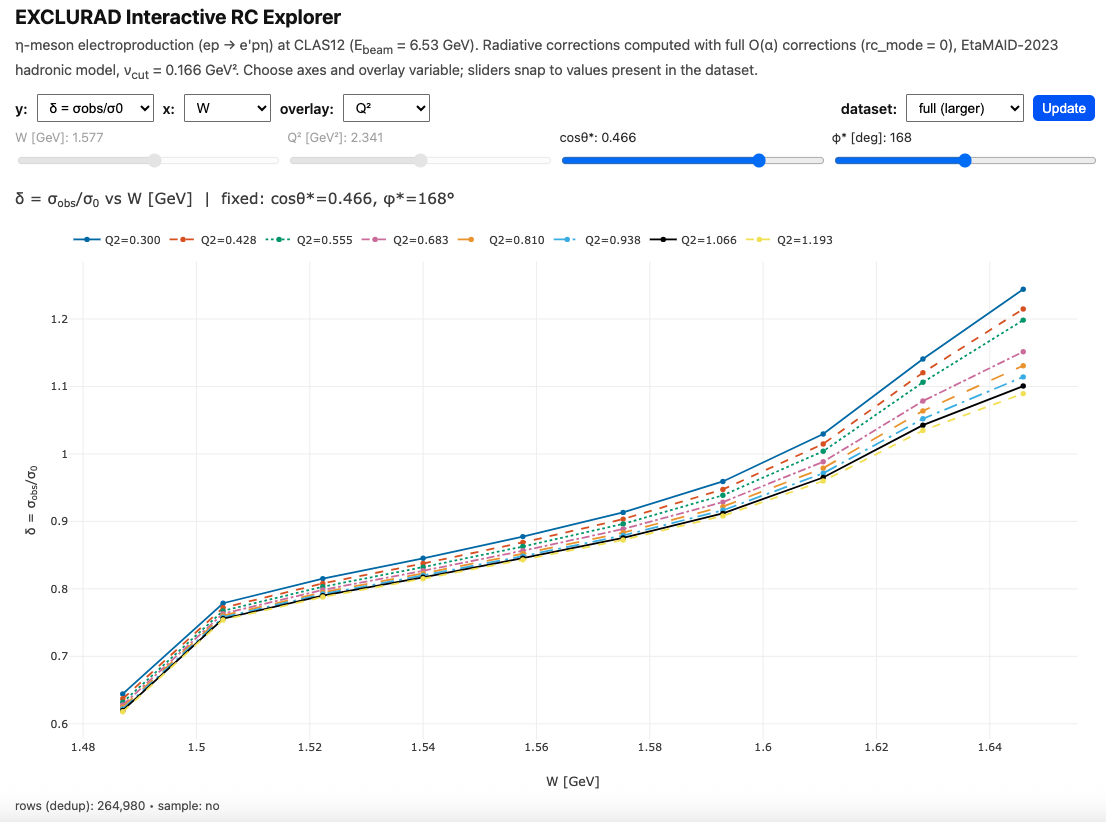}
    \caption{The EXCLURAD RC Explorer (\url{https://izzyillari.github.io/exclurad/}), showing $\delta \equiv \sigma_{\rm obs}/\sigma_0$ vs.\ $W$ with $Q^{2}$ as the overlay, at fixed $\cos\theta^{\ast} = 0.466$, $\phi^{\ast} = 168^{\circ}$. Eight $Q^{2}$ values from $0.300$ to $1.193~\mathrm{GeV}^{2}$ are shown. Full dataset (264\,980 rows); $v_{\mathrm{cut}} = 0.166~\mathrm{GeV}^{2}$.}
    \label{fig:s6_explorer}
\end{figure*}

% ------------------------------------------------------------------
A web page at \url{https://izzyillari.github.io/exclurad/} (Figure ~\ref{fig:s6_explorer}) plots $\delta$ and $R_A$ from the full EXCLURAD output across the four-dimensional $(W, Q^{2}, \cos\theta^{\ast}, \phi^{\ast})$ grid. The page is static: it reads from two Apache Arrow Feather files~\cite{ApacheArrow} shipped with the repository and renders via Plotly.js~\cite{Plotlyjs}; no server is required. The user selects $y$, $x$, and overlay variables from dropdown menus and sets the remaining two kinematic variables with sliders. Slider positions snap to grid values present in the dataset, so no interpolation is applied. Up to eight overlay curves are drawn using the Okabe-Ito palette~\cite{OkabeIto} with distinct line styles. Two dataset sizes are available: 264\,980 rows (full) and 50\,000 rows (faster load). The dataset was computed at the parameters in Table~\ref{tab:s6_params}.

\begin{table}[!htbp]
    \centering
    \caption{Parameters for the dataset served by the RC Explorer.}
    \label{tab:s6_params}
    \small
    \begin{tabular}{ll}
        \hline\hline
        \textbf{Parameter} & \textbf{Value} \\
        \hline
        Reaction              & $ep \to e'p\eta$ \\
        Beam energy           & $6.53~\mathrm{GeV}$ (CLAS12) \\
        RC mode               & Full $O(\alpha)$ \\
        Hadronic model        & EtaMAID-2023 \\
        $v_{\mathrm{cut}}$    & $0.166~\mathrm{GeV}^{2}$ \\
        $W$ range             & $1.487$--$1.999~\mathrm{GeV}$ \\
        $Q^{2}$ range         & $0.30$--$4.00~\mathrm{GeV}^{2}$ \\
        $\cos\theta^{\ast}$   & $-0.900$ to $+0.900$ \\
        $\phi^{\ast}$         & $0^{\circ}$--$360^{\circ}$, 30 bin-centers \\
        \hline\hline
    \end{tabular}
\end{table}

% ------------------------------------------------------------------
\subsubsection*{\texorpdfstring{S4.3\quad Code repository and data availability}{S4.3 Code repository and data availability}}
% ------------------------------------------------------------------
The modified EXCLURAD source, EtaMAID-2023 lookup table, run scripts, and web explorer are available at \url{https://github.com/IzzyIllari/exclurad} and archived at DOI:~\href{https://doi.org/10.5281/zenodo.18970108}{10.5281/zenodo.18970108} (v1.0.0). The pion-channel EXCLURAD from Ref.~\cite{Afanasev:2002ee} is maintained at \url{https://github.com/JeffersonLab/exclurad}. SHA-256 checksums for the primary data files are:
\begin{verbatim}
75403d64a53af1e0b7f3fb0d6daf50310c78d809e250c4f1875f1ca900ab07b6
    exclurad/maid07-PPpi.tbl
231839a8e98e9494e8bf19e5439d3b6ee04749ce4dd9571a477910f4dc35c91e
    docs/data/exclurad_eta_web.feather
a14406f84660e59026739a72da2fe1851bbb1e8109d710aa52b7687cf8afad8c
    docs/data/exclurad_eta_web_sample.feather
\end{verbatim}
%============================================================
\end{document}